\input harvmac
\input epsf
\noblackbox

\newcount\figno

\figno=0
\def\fig#1#2#3{
\par\begingroup\parindent=0pt\leftskip=1cm\rightskip=1cm\parindent=0pt
\baselineskip=11pt \global\advance\figno by 1 \midinsert
\epsfxsize=#3 \centerline{\epsfbox{#2}} \vskip 12pt
\centerline{{\bf Figure \the\figno :}{\it ~~ #1}}\par
\endinsert\endgroup\par}
\def\figlabel#1{\xdef#1{\the\figno}}
\def\pano{\par\noindent}

\font\cmss=cmss10
\font\cmsss=cmss10 at 7pt

\def\rlx{\relax\leavevmode}
\def\inbar{\vrule height1.5ex width.4pt depth0pt}
\def\IC{\relax\,\hbox{$\inbar\kern-.3em{\rm C}$}}
\def\IR{\relax{\rm I\kern-.18em R}}
\def\IN{\relax{\rm I\kern-.18em N}}
\def\IP{\relax{\rm I\kern-.18em P}}
\def\ZZ{\rlx\leavevmode\ifmmode\mathchoice{\hbox{\cmss Z\kern-.4em Z}}
 {\hbox{\cmss Z\kern-.4em Z}}{\lower.9pt\hbox{\cmsss Z\kern-.36em Z}}
 {\lower1.2pt\hbox{\cmsss Z\kern-.36em Z}}\else{\cmss Z\kern-.4em Z}\fi}

\def\narrowplus{\kern -.04truein + \kern -.03truein}
\def\narrowminus{- \kern -.04truein}
\def\narrowminussub{\kern -.02truein - \kern -.01truein}

\def\o#1{\overline{#1}}
\def\la{\langle}
\def\ra{\rangle}



\lref\GepnerVZ{
D.~Gepner,
``Exactly Solvable String Compactifications On Manifolds Of SU(N) Holonomy,''
Phys.\ Lett.\ B {\bf 199}, 380 (1987).
}

\lref\GepnerQI{
D.~Gepner,
``Space-Time Supersymmetry In Compactified String Theory And Superconformal Models,''
Nucl.\ Phys.\ B {\bf 296}, 757 (1988).
}

\lref\OoguriCK{
H.~Ooguri, Y.~Oz and Z.~Yin,
``D-branes on Calabi-Yau spaces and their mirrors,''
Nucl.\ Phys.\ B {\bf 477}, 407 (1996)
[arXiv:hep-th/9606112].
}

\lref\RecknagelSB{
A.~Recknagel and V.~Schomerus,
``D-branes in Gepner models,''
Nucl.\ Phys.\ B {\bf 531}, 185 (1998)
[arXiv:hep-th/9712186].
}

\lref\review{
A.~M.~Uranga,
``Chiral four-dimensional string compactifications with intersecting  D-branes,''
Class.\ Quant.\ Grav.\  {\bf 20}, S373 (2003), [arXiv:hep-th/0301032]\semi
F.~G.~Marchesano Buznego,
``Intersecting D-brane models,''
arXiv:hep-th/0307252\semi
T.~Ott,
``Aspects of stability and phenomenology in type IIA orientifolds with 
intersecting D6-branes,'',arXiv:hep-th/0309107.
}

\lref\BrunnerNK{
M.~Gutperle and Y.~Satoh,
``D-branes in Gepner models and supersymmetry,''
Nucl.\ Phys.\ B {\bf 543}, 73 (1999)
[arXiv:hep-th/9808080]\semi
S.~Govindarajan, T.~Jayaraman and T.~Sarkar,
``Worldsheet approaches to D-branes on supersymmetric cycles,''
Nucl.\ Phys.\ B {\bf 580}, 519 (2000)
[arXiv:hep-th/9907131]\semi
D.~E.~Diaconescu and C.~R\"omelsberger,
``D-branes and bundles on elliptic fibrations,''
Nucl.\ Phys.\ B {\bf 574}, 245 (2000)
[arXiv:hep-th/9910172]\semi
P.~Kaste, W.~Lerche, C.~A.~Lutken and J.~Walcher,
``D-branes on K3-fibrations,''
Nucl.\ Phys.\ B {\bf 582}, 203 (2000)
[arXiv:hep-th/9912147]\semi
E.~Scheidegger,
``D-branes on some one- and two-parameter Calabi-Yau hypersurfaces,''
JHEP {\bf 0004}, 003 (2000)
[arXiv:hep-th/9912188]\semi
I.~Brunner and V.~Schomerus,
``D-branes at singular curves of Calabi-Yau compactifications,''
JHEP {\bf 0004}, 020 (2000)
[arXiv:hep-th/0001132]\semi
M.~Naka and M.~Nozaki,
``Boundary states in Gepner models,''
JHEP {\bf 0005}, 027 (2000)
[arXiv:hep-th/0001037].
}

\lref\BrunnerJQ{
I.~Brunner, M.~R.~Douglas, A.~E.~Lawrence and C.~R\"omelsberger,
``D-branes on the quintic,''
JHEP {\bf 0008}, 015 (2000)
[arXiv:hep-th/9906200].
}

\lref\MisraZV{
A.~Misra,
``On (orientifold of) type IIA on a compact Calabi-Yau,''
arXiv:hep-th/0304209.
}

\lref\MizoguchiXI{
S.~Mizoguchi and T.~Tani,
``Wound D-branes in Gepner models,''
Nucl.\ Phys.\ B {\bf 611}, 253 (2001)
[arXiv:hep-th/0105174].
}

\lref\GovindarajanJS{
S.~Govindarajan, T.~Jayaraman and T.~Sarkar,
``Worldsheet approaches to D-branes on supersymmetric cycles,''
Nucl.\ Phys.\ B {\bf 580}, 519 (2000)
[arXiv:hep-th/9907131].
}

\lref\BrunnerZM{
I.~Brunner and K.~Hori,
``Orientifolds and mirror symmetry,''
arXiv:hep-th/0303135.
}

\lref\BrunnerEM{
I.~Brunner and K.~Hori,
``Notes on orientifolds of rational conformal field theories,''
[arXiv:hep-th/0208141].
}

\lref\BrunnerFS{
I.~Brunner,
``On orientifolds of WZW models and their relation to geometry,''
JHEP {\bf 0201}, 007 (2002)
[arXiv:hep-th/0110219].
}

\lref\HuiszoonAI{
L.~R.~Huiszoon and K.~Schalm,
``BPS orientifold planes from crosscap states in Calabi-Yau  compactifications,''
[arXiv:hep-th/0306091].
}

\lref\HuiszoonVY{
L.~R.~Huiszoon,
``Comments on the classification of orientifolds,''
Class.\ Quant.\ Grav.\  {\bf 20}, S509 (2003)
[arXiv:hep-th/0212244].
}

\lref\FuchsCM{
J.~Fuchs, L.~R.~Huiszoon, A.~N.~Schellekens, C.~Schweigert and J.~Walcher,
``Boundaries, crosscaps and simple currents,''
Phys.\ Lett.\ B {\bf 495}, 427 (2000)
[arXiv:hep-th/0007174].
}

\lref\HuiszoonGE{
L.~R.~Huiszoon and A.~N.~Schellekens,
``Crosscaps, boundaries and T-duality,''
Nucl.\ Phys.\ B {\bf 584}, 705 (2000)
[arXiv:hep-th/0004100].
}

\lref\HuiszoonXQ{
L.~R.~Huiszoon, A.~N.~Schellekens and N.~Sousa,
``Klein bottles and simple currents,''
Phys.\ Lett.\ B {\bf 470}, 95 (1999)
[arXiv:hep-th/9909114].
}

\lref\GovindarajanVP{
S.~Govindarajan and J.~Majumder,
``Crosscaps in Gepner models and type IIA orientifolds,''
[arXiv:hep-th/0306257].
}

\lref\GovindarajanVV{
S.~Govindarajan and J.~Majumder,
``Orientifolds of type IIA strings on Calabi-Yau manifolds,''
arXiv:hep-th/0305108.
}

\lref\BlumenhagenJB{
R.~Blumenhagen and V.~Braun,
``Superconformal field theories for compact G(2) manifolds,''
JHEP {\bf 0112}, 006 (2001)
[arXiv:hep-th/0110232]\semi
R.~Roiban and J.~Walcher,
``Rational conformal field theories with G(2) holonomy,''
JHEP {\bf 0112}, 008 (2001)
[arXiv:hep-th/0110302]\semi
T.~Eguchi and Y.~Sugawara,
``String theory on G(2) manifolds based on Gepner construction,''
Nucl.\ Phys.\ B {\bf 630}, 132 (2002)
[arXiv:hep-th/0111012].
}

\lref\BlumenhagenJY{
R.~Blumenhagen, D.~L\"ust and S.~Stieberger,
``Gauge unification in supersymmetric intersecting brane worlds,''
JHEP {\bf 0307}, 036 (2003)
[arXiv:hep-th/0305146].
}

\lref\BlumenhagenMD{
R.~Blumenhagen, L.~G\"orlich and B.~K\"ors,
``Supersymmetric orientifolds in 6D with D-branes at angles,''
Nucl.\ Phys.\ B {\bf 569}, 209 (2000)
[arXiv:hep-th/9908130]\semi
R.~Blumenhagen, L.~G\"orlich and B.~K\"ors,
``Supersymmetric 4D orientifolds of type IIA with D6-branes at angles,''
JHEP {\bf 0001}, 040 (2000)
[arXiv:hep-th/9912204]\semi
G.~Pradisi,
``Type I vacua from diagonal Z(3)-orbifolds,''
Nucl.\ Phys.\ B {\bf 575}, 134 (2000)
[arXiv:hep-th/9912218]\semi
S.~F\"orste, G.~Honecker and R.~Schreyer,
``Supersymmetric Z(N) x Z(M) orientifolds in 4D with D-branes at angles,''
Nucl.\ Phys.\ B {\bf 593}, 127 (2001)
[arXiv:hep-th/0008250].
}

\lref\BlumenhagenTJ{
R.~Blumenhagen and A.~Wisskirchen,
``Spectra of 4D, N = 1 type I string vacua on non-toroidal CY threefolds,''
Phys.\ Lett.\ B {\bf 438}, 52 (1998)
[arXiv:hep-th/9806131].
}

\lref\AngelantonjMW{
C.~Angelantonj, M.~Bianchi, G.~Pradisi, A.~Sagnotti and Y.~S.~Stanev,
``Comments on Gepner models and type I vacua in string theory,''
Phys.\ Lett.\ B {\bf 387}, 743 (1996)
[arXiv:hep-th/9607229].
}

\lref\AldazabalUB{
G.~Aldazabal, E.~C.~Andres, M.~Leston and C.~Nunez,
``Type IIB orientifolds on Gepner points,''
JHEP {\bf 0309}, 067 (2003)
[arXiv:hep-th/0307183].
}

\lref\FuchsGV{
J.~Fuchs, C.~Schweigert and J.~Walcher,
``Projections in string theory and boundary states for Gepner models,''
Nucl.\ Phys.\ B {\bf 588}, 110 (2000)
[arXiv:hep-th/0003298].
}

\lref\RecknagelQQ{
A.~Recknagel,
``Permutation branes,''
JHEP {\bf 0304}, 041 (2003)
[arXiv:hep-th/0208119].
}

\lref\stanev{
Y.~S.~Stanev,''Open Descendants of Gepner Models in 6D'', talk presented at the
workshop ``Conformal Field Theory of D=branes'', at DESY, Hamburg, Germany,
Transparancies on http://www.desy.de/$\sim$jfuchs/CftD.html.
}

\lref\rangles{M.~Berkooz, M.R.~Douglas and R.G.~Leigh, {\it Branes Intersecting
at Angles}, Nucl. Phys. B {\bf 480} (1996) 265, hep-th/9606139.
}

\lref\MaldacenaKY{
J.~M.~Maldacena, G.~W.~Moore and N.~Seiberg,
``Geometrical interpretation of D-branes in gauged WZW models,''
JHEP {\bf 0107}, 046 (2001)
[arXiv:hep-th/0105038].
}

\lref\rbgklnon{R.~Blumenhagen, L.~G\"orlich, B.~K\"ors and D.~L\"ust,
``Noncommutative Compactifications of Type I Strings on Tori with Magnetic
Background Flux'', JHEP {\bf 0010}, 006 (2000), [arXiv:hep-th/0007024] \semi
R.~Blumenhagen, B.~K\"ors and D.~L\"ust,
``Type I Strings with $F$ and $B$-Flux'', JHEP {\bf 0102}, 030 (2001),
[arXiv:hep-th/0012156] \semi
R.Blumenhagen, B.~K\"ors, D.~L\"ust and T.~Ott,
``The standard model from stable intersecting brane world orbifolds'',
Nucl.\ Phys.\ B {\bf 616}, 3 (2001)
[arXiv:hep-th/0107138].
}

\lref\raads{C.~Angelantonj, I.~Antoniadis, E.~Dudas, A.~Sagnotti, ``Type I
Strings on Magnetized Orbifolds and Brane Transmutation'',
Phys. Lett. B {\bf 489}, 223 (2000), [arXiv:hep-th/0007090]\semi
C.~Angelantonj, A.~Sagnotti, ``Type I
Vacua and Brane Transmutation'', [arXiv:hep-th/0010279].
}

\lref\timo{R.~Blumenhagen and T.~Weigand. {\it in preparation}.}

\lref\rbbkl{R.~Blumenhagen, V.~Braun, B.~K\"ors and D.~L\"ust,
``Orientifolds of K3 and Calabi-Yau Manifolds with Intersecting D-branes'',
JHEP {\bf 0207}, 026  (2002), [arXiv:hep-th/0206038]\semi
R.~Blumenhagen, V.~Braun, B.~K\"ors and D.~L\"ust,
``The standard model on the quintic,''
arXiv:hep-th/0210083.
}

\lref\rcveticb{M.~Cvetic, G.~Shiu and  A.M.~Uranga,  ``Three-Family
Supersymmetric Standard-like Models from Intersecting Brane Worlds'',
Phys. Rev. Lett. {\bf 87},  201801 (2001),  [arXiv:hep-th/0107143]\semi
M.~Cvetic, G.~Shiu and  A.M.~Uranga,  
``Chiral Four-Dimensional N=1 Supersymmetric Type IIA Orientifolds from
Intersecting D6-Branes'', Nucl. Phys. B {\bf 615}, 3  (2001), [arXiv:hep-th/0107166]\semi
R.~Blumenhagen, L.~G\"orlich and T.~Ott,
``Supersymmetric intersecting branes on the type IIA 
$T^6/Z(4)$  orientifold'', JHEP {\bf 0301}, 021 (2003),
[arXiv:hep-th/0211059]\semi
G.~Honecker,
``Chiral supersymmetric models on an orientifold of Z(4) x Z(2) with 
 intersecting D6-branes,''
Nucl.\ Phys.\ B {\bf 666}, 175 (2003)
[arXiv:hep-th/0303015]\semi
M.~Larosa and G.~Pradisi,
``Magnetized four-dimensional Z(2) x Z(2) orientifolds,''
Nucl.\ Phys.\ B {\bf 667}, 261 (2003)
[arXiv:hep-th/0305224].
}

\lref\rbachas{C.~Bachas, ``A Way to Break Supersymmetry'', [arXiv:hep-th/9503030]\semi 
M.~Berkooz, M.R.~Douglas and R.G.~Leigh, ``Branes Intersecting
at Angles'', Nucl. Phys. B {\bf 480}, 265  (1996), [arXiv:hep-th/9606139].
}

\lref\rafiru{G.~Aldazabal, S.~Franco, L.E.~Ibanez, R.~Rabadan, A.M.~Uranga,
{\it $D=4$ Chiral String Compactifications from Intersecting Branes},
J.\ Math.\ Phys.\  {\bf 42}, 3103 (2001), [arXiv:hep-th/0011073]\semi
G.~Aldazabal, S.~Franco, L.E.~Ibanez, R.~Rabadan, A.M.~Uranga,
{\it Intersecting Brane Worlds}, JHEP {\bf 0102}, 047 (2001), [arXiv:hep-ph/0011132]\semi
L.E.~Ibanez, F.~Marchesano, R.~Rabadan, {\it Getting just the
Standard Model at Intersecting Branes},
JHEP {\bf 0111}, 002 (2001), [arXiv:hep-th/0105155].
}

\lref\UrangaPZ{
A.~M.~Uranga,
``Chiral four-dimensional string compactifications with intersecting  D-branes,''
Class.\ Quant.\ Grav.\  {\bf 20}, S373 (2003)
[arXiv:hep-th/0301032].
}

\lref\UrangaPG{
A.~M.~Uranga,
``Local models for intersecting brane worlds,''
JHEP {\bf 0212}, 058 (2002)
[arXiv:hep-th/0208014].
}

\lref\BianchiYU{
M.~Bianchi and A.~Sagnotti,
``On The Systematics Of Open String Theories,''
Phys.\ Lett.\ B {\bf 247}, 517 (1990)\semi
M.~Bianchi and A.~Sagnotti,
``Twist Symmetry And Open String Wilson Lines,''
Nucl.\ Phys.\ B {\bf 361}, 519 (1991)\semi
M.~Bianchi, G.~Pradisi and A.~Sagnotti,
``Toroidal compactification and symmetry breaking in open string theories,''
Nucl.\ Phys.\ B {\bf 376}, 365 (1992).
}

\lref\AngelantonjCT{
C.~Angelantonj and A.~Sagnotti,
``Open strings,''
Phys.\ Rept.\  {\bf 371}, 1 (2002)
[Erratum-ibid.\  {\bf 376}, 339 (2003)]
[arXiv:hep-th/0204089].
}

\lref\PradisiQY{
G.~Pradisi, A.~Sagnotti and Y.~S.~Stanev,
``Planar duality in SU(2) WZW models,''
Phys.\ Lett.\ B {\bf 354}, 279 (1995)
[arXiv:hep-th/9503207]\semi
G.~Pradisi, A.~Sagnotti and Y.~S.~Stanev,
``The Open descendants of nondiagonal SU(2) WZW models,''
Phys.\ Lett.\ B {\bf 356}, 230 (1995)
[arXiv:hep-th/9506014]\semi
G.~Pradisi, A.~Sagnotti and Y.~S.~Stanev,
``Completeness Conditions for Boundary Operators in 2D Conformal Field Theory,''
Phys.\ Lett.\ B {\bf 381}, 97 (1996)
[arXiv:hep-th/9603097].
}

\lref\CallanPX{
C.~G.~Callan, C.~Lovelace, C.~R.~Nappi and S.~A.~Yost,
``Adding Holes And Crosscaps To The Superstring,''
Nucl.\ Phys.\ B {\bf 293}, 83 (1987)\semi
J.~Polchinski and Y.~Cai,
``Consistency Of Open Superstring Theories,''
Nucl.\ Phys.\ B {\bf 296}, 91 (1988).

}

\lref\GimonAY{
E.~G.~Gimon and J.~Polchinski,
``Consistency Conditions for Orientifolds and D-Manifolds,''
Phys.\ Rev.\ D {\bf 54}, 1667 (1996)
[arXiv:hep-th/9601038]\semi
E.~G.~Gimon and C.~V.~Johnson,
``K3 Orientifolds,''
Nucl.\ Phys.\ B {\bf 477}, 715 (1996)
[arXiv:hep-th/9604129].
}

\lref\BerkoozDW{
M.~Berkooz and R.~G.~Leigh,
``A D = 4 N = 1 orbifold of type I strings,''
Nucl.\ Phys.\ B {\bf 483}, 187 (1997)
[arXiv:hep-th/9605049]\semi
C.~Angelantonj, M.~Bianchi, G.~Pradisi, A.~Sagnotti and Y.~S.~Stanev,
``Chiral asymmetry in four-dimensional open- string vacua,''
Phys.\ Lett.\ B {\bf 385}, 96 (1996)
[arXiv:hep-th/9606169]\semi
Z.~Kakushadze,
``Aspects of N = 1 type I-heterotic duality in four dimensions,''
Nucl.\ Phys.\ B {\bf 512}, 221 (1998)
[arXiv:hep-th/9704059]\semi
Z.~Kakushadze and G.~Shiu,
``A chiral N = 1 type I vacuum in four dimensions and its heterotic dual,''
Phys.\ Rev.\ D {\bf 56}, 3686 (1997)
[arXiv:hep-th/9705163]\semi
G.~Aldazabal, A.~Font, L.~E.~Ibanez and G.~Violero,
``D = 4, N = 1, type IIB orientifolds,
Nucl.\ Phys.\ B {\bf 536}, 29 (1998)
[arXiv:hep-th/9804026].
}

\Title{\vbox{
 \hbox{DAMTP-2003-112}
 \hbox{hep-th/0310244}}}
{\vbox{\centerline{ Supersymmetric Orientifolds of Gepner Models} 
}}
\centerline{Ralph Blumenhagen{} }
\bigskip\medskip
\centerline{ {\it DAMTP, Centre for Mathematical Sciences,}}
\centerline{\it Wilberforce Road, Cambridge CB3 0WA, UK}
\centerline{\tt Email:R.Blumenhagen@damtp.cam.ac.uk}
\bigskip
\bigskip

\centerline{\bf Abstract}
\noindent
Supersymmetric orientifolds of four dimensional Gepner Models
are constructed in a systematic way.
For all levels of the Gepner model  being odd the generic expression
for both the A-type and the B-type Klein bottle amplitude is derived. 
The appearing massless tadpoles are canceled 
by introducing appropriate boundary states of Recknagel/Schomerus(RS). 
After determining the M\"obius strip amplitude
we extract general expressions for the tadpole cancellation conditions. 
We discuss the issue of chirality for such
supersymmetric orientifold models and finally present a couple
of examples in detail.


\Date{10/2003}
\newsec{Introduction}

The last years have seen some encouraging progress in string model building
as well as in understanding D-branes in curved space-times.  
In particular we have learned that intersecting D-brane models
\refs{\rbachas\rbgklnon\raads\rafiru\rcveticb-\UrangaPZ}
do provide phenomenologically appealing string vacua, where
many aspects of the Standard Model, like chirality or
family replication,  can be achieved quite naturally (for reviews see \review).
For instance, the number of families is given by the
topological intersection number between pairs of D-branes
wrapping some 3-cycles in the internal space.  
However, it turned out that quasi realistic 
supersymmetric models, in particular those with no adjoint
scalars, are much more 
difficult to get, at least in the toroidal orbifold setting
one has used so far \rcveticb. Therefore, from this point of view it is 
necessary to move beyond the simple toroidal orbifold backgrounds.

One step in this direction was performed in \rbbkl, where the structure
of  intersecting  D-brane models on general Calabi-Yau spaces
was discussed (see also \refs{\UrangaPG,\MisraZV}). Unfortunately, for general Calabi-Yau spaces
not very much is known
about the special Lagrangian 3-cycles the D6-branes are supposed
to wrap around, hampering any further progress 
in finding any  concrete realization of such intersecting brane
models. 

However, it is known that certain points in the moduli
space of Calabi-Yau compactifications are exactly solvable and that 
their concrete rational conformal field theory is described
by  so-called Gepner models \refs{\GepnerQI,\GepnerVZ}. 
After same more general discussion in \OoguriCK, triggered by  the pioneering work of
Recknagel/Schomerus \RecknagelSB\   there has been some amount of study on
boundary states in these Gepner models 
\refs{\BrunnerNK\BrunnerJQ\FuchsGV\MizoguchiXI-\RecknagelQQ}. 
Geometrically these boundary states correspond to 
stable supersymmetric D-branes either wrapping special Lagrangian
3-cycles in the Calabi-Yau  or wrapping holomorphic curves equipped
with some coherent sheaves. Part of the data one needs for constructing
intersecting D-brane models, like the intersection numbers, are
known for these boundary states \BrunnerJQ. Therefore, it is a natural
question whether one can actually use them in a concrete way 
for model building.

Clearly for finally yielding supersymmetric backgrounds
it is mandatory to cancel the positive tension of these
branes by same negative contributions arising naturally
from orientifolds planes. 
Therefore,  
we are naturally led to consider orientifolds  of Gepner models 
\refs{\AngelantonjMW\stanev\BlumenhagenTJ\BrunnerEM\BrunnerZM\GovindarajanVV
\HuiszoonAI\GovindarajanVP-\AldazabalUB}. 
We would like to emphasize that even without these phenomenological ambitions in mind, 
just from the
technical conformal field theory  point of view 
\refs{\BianchiYU\PradisiQY\HuiszoonXQ\HuiszoonGE\FuchsCM\BrunnerFS-\HuiszoonVY}, 
the construction of fully fledged Gepner model orientifolds  
is a natural next step to perform.

In fact there has lately been some work on the construction
of crosscap states (orientifold planes) in Gepner models
respectively in more general $N=2$ theories 
\refs{\BrunnerZM,\GovindarajanVV,
\HuiszoonAI,\GovindarajanVP}. However, there
do exist only very few attempts to really construct fully
fledged orientifold models including the derivation
of the tadpole cancellation conditions, their solutions
and the determination of the resulting massless spectra.
After constructing some six dimensional models in \AngelantonjMW, the
first four dimensional models were studied in \BlumenhagenTJ, where it came
as a surprise that the gauge groups could have a much larger rank
than  initially expected. For instance, it turned out that for
the $(3)^5$ Gepner model the gauge group could maximally be 
$G=SO(12)\times SO(20)$. Recently, in \AldazabalUB\ many more
eight and six-dimensional models were constructed in addition
to the example of the four-dimensional $(3)^5$ Gepner model.

So far the construction of models were more like a case by case study
with intensive use of computer power. 
An important step forward would be to derive general  
results for the Klein bottle and M\"obius amplitudes
or for the tadpole
cancellation conditions.
The aim of the current paper is to approach this problem and start
a systematic investigation of orientifolds of Gepner model.
In the course of the paper
we will consider four dimensional B-type and A-type 
orientifolds of Type IIB on Gepner models.
To avoid the subtlety of simple current
fixed points and to keep the formulas as simple as possible, 
we restrict ourselves to the case where all
levels of the Gepner model are odd. 

We will show that it is indeed possible to derive absolutely  general
expressions for the Klein bottle amplitude. 
The appearing massless tadpoles are canceled by introducing RS-boundary
states, which pairwise generically break supersymmetry and 
have non-zero intersection numbers meaning that there
are   chiral fermions localized on the intersection locus. 
In this respect they are completely 
analogous to the D-branes intersecting at angles, which were 
introduced to get chiral fermions in toroidal orbifold models. 
However, it will turn out that the requirement of supersymmetry further 
constrains the model and that relatively supersymmetric B-type boundary states do not admit 
chiral fermions anymore. 

The most sophisticated step is the determination of the M\"obius amplitude,
which involves extra sign factors in front of the characters. 
Led by the requirement that the loop and tree channel  M\"obius amplitudes
do respect the GSO projection, we will determine these signs and compute
(for simplicity just in the NS sector) general expressions for
them. Adding up all different one-loop amplitudes we
will extract the massless tadpole cancellation conditions, which turn
out to have a very suggestive form. All these amplitudes and 
conditions can be straightforwardly put on a computer to search 
for interesting non-trivial solutions. 
We will discuss a couple of simple B-type examples and leave the
more complicated though probably phenomenologically more appealing
 A-type models for future study. 

This paper is organized as follows. In section 2 we review some
facts about Gepner models. Section 3 contains some comments
about orientifolds of Gepner models in general followed by the computation
of the Klein bottle amplitude for the B-type and A-type orientifolds. 
In section 4 we review some of the important aspects of 
RS-boundary states including the loop and tree channel
annulus amplitudes. The derivation of the M\"obius strip
amplitudes is the subject of section 5, where for simplicity
we restrict ourselves to the explicit computation of the 
NS sector amplitudes. In section 6 we present the general tadpole
cancellation conditions and discuss a couple of examples
in section 7. Finally, section 8 contains our conclusions.

\newsec{Review of Gepner Models}

To set the stage we briefly review some aspects of Gepner models
needed in the remainder of this paper. 
In light cone gauge, the internal sector of a Type II compactification
to four dimensions with N=2 supersymmetry is given
by an N=2 supersymmetric conformal field theory (SCFT) with central
charge $c=9$. The idea of Gepner was to use tensor products of 
the well known rational models of the N=2 super Virasoro algebra 
for this N=2 SCFT \refs{\GepnerQI,\GepnerVZ}. 
In order to really get space-time supersymmetry
one has to invoke a GSO projection, which can be described
by a certain simple current in the SCFT.

More concretely, the minimal models are parametrized by the so-called
level $k=1,2,\ldots$, in term of which the central charge
is given by
\eqn\num{
c={3k\over k+2} .}
Since $c<3$ one has to use tensor products of such minimal models
$ \bigotimes_{j=1}^r (k_j)$ adding  up to the required value $c=9$. 
Each unitary model has only a finite number
of irreducible representations of the N=2 Virasoro algebra, which 
are labeled by the three integers $(l,m,s)$ in the range
\eqn\range{  l=0,\ldots k, \quad m=-k-1,-k,\ldots k+2, \quad
              s=-1,0,1,2 }
with $l+m+s=0$ mod $2$. 
Actually, taking the identification between $(l,m,s)$ and
$(k-l,m+k+2,s+2)$ into account, the range \range\  is a double covering of the
allowed representations. 
Since the Cartan subalgebra of the N=2 Virasoro algebra contains
the two elements $L_0,j_0$, each state in the representation
carries two quantum numbers, the conformal weight and
the $U(1)$ charge. 
Utilizing the coset construction of each minimal model as 
$SU(2)_k\times U(1)_2/U(1)_{k+2}$,
the conformal dimension $h$ and charge $q$ of the highest weight 
state with label $(l,m,s)$ is given by
\eqn\quannum{\eqalign{
\Delta^l_{m,s}&={l(l+2)-m^2\over 4(k+2)} + {s^2\over 8}\quad ({\rm mod}\ 1) \cr
q^l_{m,s}&={m\over (k+2)}-{s\over 2}\quad ({\rm mod}\ 2) .}}
Note, that these formulas are  only correct modulo one respectively two. 
The exact conformal dimension $h$ and charge can be read off from \quannum\
by  first shifting  the labels into the standard range $|m-s|\le l$
by using the shift symmetries $m\to m+2k+4$,$s\to s+4$ and 
the reflection symmetry.  
Representations with even $s$ belong to the NS-sector while those
with odd $s$ belong to the R-sector. 

So far we have just given the field content of each $N=2$ tensor factor.
In addition one has the contributions with $c=3$ from the two uncompactified
directions where the two world-sheet fermions $\psi^{2,3}$ 
generate a $U(1)_2$ model. This has four irreducible
representations labeled by $s_0=-1,\ldots 2$ with highest weight
and charge 
\eqn\fermi{ \Delta_{s_0}={s_0^2\over 8}  \ \  ({\rm mod}\ 1),
             \quad\quad q_{s_0}=-{s_0\over 2}\quad ({\rm mod}\ 2 ).}
In the superstring construction in order to achieve space-time
supersymmetry one has to implement a GSO projection, which in
the Gepner case means that one projects onto states 
with odd overall $U(1)$ charge $Q_{tot}=q_{s_0}+\sum_{j=1}^r q^{l_j}_{m_j,s_j}$.
Moreover, to have a good space-time interpretation one has to ensure
that in the tensor product only states from the NS respectively 
the R sectors couple among themselves.

In order to describe these projections in a simple way one
introduces the following notations.
First one defines some multi-labels
\eqn\mlabels{  \lambda=(l_1,\ldots,l_r), \quad \mu=(s_0;m_1,\ldots m_r; 
                s_1,\ldots,s_r) }
and the respective characters 
\eqn\mchar{ \chi^\lambda_\mu(q)=\chi_{s_0}(q)\, \chi^{l_1}_{m_1,s_1}(q)
                        \ldots \chi^{l_r}_{m_r,s_r}(q) .}
Introducing the vectors 
\eqn\mvec{ \beta_0=(1;1,\ldots,1;1,\ldots,1), \quad 
            \beta_j=(2;0,\ldots,0;0,\ldots,0,\underbrace{2}_{j^{\rm th}},0,\ldots,0) }
and the following product
\eqn\mprod{\eqalign{  Q_{tot}&=2\beta_0\bullet \mu =-{s_0\over 2}-\sum_{j=1}^r{s_j\over 2}
              +\sum_{j=1}^r {m_j\over k_j+2},\cr
                  \beta_j\bullet \mu &=-{s_0\over 2}-{s_j\over 2},   }}
the projections one has to implement are simply 
$Q_{tot}=2\beta_0\bullet \mu\in 2\ZZ+1$ and $\beta_j\bullet \mu \in \ZZ$
for all $j=1,\ldots r$.
Gepner has shown that the following GSO projected
partition function
\eqn\parti{    Z_D(\tau,\o{\tau} )={1\over 2^r} 
{ ({\rm Im}\tau)^{-2} \over |\eta(q)|^4 }
     \sum_{b_0=0}^{K-1} \sum_{b_1,\ldots,b_r=0}^1 {\sum_{\lambda,\mu}}^\beta
    (-1)^{s_0} \ \chi^\lambda_\mu (q)\, \chi^\lambda_{\mu+b_0\beta_0
              +b_1 \beta_1 +\ldots b_r \, \beta_r} (\o q) }
is indeed modular invariant and vanishes due to space-time supersymmetry. 
Here $K={\rm lcm}(4,2k_j+4)$ and ${\sum}^\beta$ means that the sum is 
restricted to 
those $\lambda$ and $\mu$ in the range \range\ satisfying 
$2\beta_0\bullet \mu\in 2\ZZ+1$ and $\beta_j\bullet \mu \in \ZZ$.
The factor $2^r$ due to  the field
identifications guarantees the correct normalization of the amplitude. 
In the partition function \parti\ states with odd charge are arranged
in orbits under the action of the $\beta$ vectors. Therefore, the partition
function is non-diagonal in the original characters, but
for all levels odd it can be written as a diagonal partition
function in terms of the orbits which in this case have
all equal length $2^r\, K$.  These
orbits under the $\beta$-vectors \mvec\ will loosely be called GSO orbits
in the following.  

Since in the sequel we will make extensive use of the modular S-transformation
rules for the characters involved in \parti, let us also state them 
here.
For the $SU(2)_k$
Kac-Moody algebra  the S-matrix  is given by
\eqn\ssmatrix{  S_{l,l'}=\sqrt{2\over k+2}\, \sin (l,l')_k ,}
where we have used the convention $(l,l')_k={\pi(l+1)(l'+1)\over k+2}$.
For the N=2 minimal model one obtains for the modular S-matrix 
\eqn\smatrix{\eqalign{    S^{U(1)_2}_{s_0,s_0'}&={1\over 2} e^{-i\pi{s_0 s'_0
        \over 2}}, \cr
        {\cal S}_{(l,m,s),(l',m',s')}&={1\over 2\sqrt{2k+4}}\, S_{l,l'}\,
        e^{i\pi{m\, m' \over k+2}}\, e^{-i\pi{s\, s' \over 2}}  .}}
Note, that in  the  latter expression there is a difference in the normalization
of the S-matrix of a factor
$\sqrt{2}$ compared to most of the literature. 
This is due to the fact that the matrix ${\cal S}^2$ only turns
out to be diagonal, if one finally identifies labels which are
related by the flip $(l,m,s)\to (k-l,m+k+2,s+2)$. 
  
As is well known in defining crosscap states one also needs
the so-called P-matrix, which is defined as 
$P=T^{1\over 2}S\,T^2\,S\,T^{1\over 2}$ and relates the loop channel
and tree channel M\"obius amplitudes.
For just the $SU(2)_k$ Kac-Moody algebra it is given by
\eqn\ppmatrix{  P_{l,l'}={2\over \sqrt{k+2}}\, 
     \sin {1\over 2}(l,l')_k\, \delta_{l+l'+k,0}^{(2)} }
and for the N=2 unitary models it reads \BrunnerZM
\eqn\pmatrix{\eqalign{    P^{U(1)_2}_{s_0,s'_0}&={1\over \sqrt{2}}\, \sigma_{s_0}\sigma_{s'_0}
  e^{-i\pi{s_0 s'_0 \over 4}}\, \delta_{s_0+s_0',0}^{(2)}, \cr
        {\cal P}_{(l,m,s),(l',m',s')}&={1\over 2\sqrt{2k+4}}\, \sigma_{(l,m,s)}\, 
   \sigma'_{(l',m',s')} \, e^{{i\pi\over 2}{m\, m' \over k+2}}\,
    e^{-i\pi{s\, s' \over 4}}\, \delta_{s+s',0}^{(2)}\cr
    &\left[
    P_{l,l'}\, \delta_{m+m'+k+2,0}^{(2)}+ (-1)^{l'+m'+s'\over 2}\,
     e^{i\pi{m+s\over 2}}\, P_{l,k-l'}\, \delta_{m+m',0}^{(2)}
   \right] .}}
The extra sign factors in \pmatrix\ are given by
\eqn\sino{\eqalign{  \sigma_{s_0}&=(-1)^{h_{s_0}-\Delta_{s_0}} \cr
    \sigma_{(l,m,s)}&=(-1)^{h^l_{m,s}-\Delta^l_{m,s}} }}
and arise, as roots of the modular $T$ matrix appear in the definition
of $P$. Compared to \BrunnerZM, for the same reason as for the matrix ${\cal S}$ 
we use a  different normalization of the matrix ${\cal P}$
\foot{Note, that for $s_0$ even one always has $\sigma_{s_0}=1$, which is one of the
reasons why the NS sector M\"obius amplitude is easier to compute in a general
way without caring about  an extensive amount of extra signs.}.

Since in the following we restrict ourselves to the case of all
levels being  odd, we present in Table 1 all Gepner models of this type
and their corresponding Calabi-Yau manifold.  
\vskip 0.8cm
\vbox{ \centerline{\vbox{ \hbox{\vbox{\offinterlineskip
\def\tablespace{height2pt&\omit&&\omit&&
 \omit&\cr}
\def\tablerule{\tablespace\noalign{\hrule}\tablespace}

\hrule\halign{&\vrule#&\strut\hskip0.2cm\hfill #\hfill\hskip0.2cm\cr
& levels && $(h_{21},h_{11})$ && CY &\cr
\tablerule
& $(1^9)$ &&  $(84,0)$ && $-$ &\cr
\tablespace
& $(1,1,3,7,43)$ &&  $(67,19)$ &&  $\IP_{1,5,9,15,15}[45]$ &\cr
\tablespace
& $(1,1,3,13,13)$ &&  $(103,7)$ &&  $\IP_{1,1,3,5,5}[15]$ &\cr
\tablespace
& $(1,1,5,5,19)$ &&  $(65,17)$ &&  $\IP_{1,3,3,7,7}[21]$ &\cr
\tablespace
& $(1,1,7,7,7)$ &&  $(112,4)$ &&  $\IP_{1,1,1,3,3}[9]$ &\cr
\tablespace
& $(1,3,3,3,13)$ &&  $(75,3)$ &&  $\IP_{1,3,3,3,5}[15]$ &\cr
\tablespace
& $(3,3,3,3,3)$ &&  $(101,1)$ &&  $\IP_{1,1,1,1,1}[5]$ &\cr
}\hrule}}}} 
\centerline{ \hbox{{\bf
Table 1:}{\it ~~ odd level Gepner models}}} } 
\vskip 0.5cm
\noindent
Apparently, for all levels odd the number of tensor factors
is either five or nine. Therefore  the formulas to be presented in
the following sections are derived under the assumption of $r=5,9$
and all levels $k_j$ odd.

\newsec{The orientifold projection}

In this section we will consider orientifolds  of the original
Gepner models. Before discussing  the Klein bottle amplitude,
we would like to make some general comments about possible
orientifold models.

\subsec{Different types of orientifolds}

The partition function \parti\ is the so called  diagonal invariant in the
sense that it combines left and right moving states with the same
$U(1)$ charge. Geometrically, the resulting model describes the
Type IIA/B string compactified on a Calabi-Yau space $M$.
For instance, for the $(3)^5$ Gepner model the corresponding 
Calabi-Yau is just the quintic hypersurface in $\IC\IP_4$. 
Besides the diagonal partition function $Z_D$ there also exists
the charge conjugated  partition function $Z_C$ where one combines
left and right moving states with opposite $U(1)$ charges in each
tensor factor. 
As is well known this describes the Type IIA/B string
on the mirror manifold $W$.  
Moreover, under mirror symmetry Type IIB with the diagonal
invariant is mapped to Type IIA with the charge conjugated
invariant and vice versa. 

These relations continue to hold if we perform orientifold
constructions, which break the space-time supersymmetry
down to $N=1$. In an orientifold one  divides out by the
world-sheet parity transformation $\Omega$ \foot{There exist more
general orientifold models where one combines $\Omega$ with some
holomorphic involution of the Calabi-Yau. However, such models
are not considered in this paper.}.  Here the $\Omega$
orientifold of Type IIB on $Z_{D/C}$ is mapped
to the $\Omega\o\sigma$ orientifold of Type IIA on $Z_{C/D}$,
where $\o\sigma$ denotes $U(1)$  charge conjugation in each
tensor factor. 
For reasons which will become clear below  we call
the Type IIB orientifold of $Z_D$ the  B-type model and
the orientifold of $Z_C$ the A-type model.
All these relations are summarized in Table 2, where entries
in the same line are related via mirror symmetry.
\vskip 0.8cm
\vbox{ \centerline{\vbox{ \hbox{\vbox{\offinterlineskip
\def\tablespace{height2pt&\omit&&\omit&&
 \omit&\cr}
\def\tablerule{\tablespace\noalign{\hrule}\tablespace}

\hrule\halign{&\vrule#&\strut\hskip0.2cm\hfill #\hfill\hskip0.2cm\cr
& && Type IIB  && Type IIA &\cr
\tablerule
& proj. &&  $\Omega$ &&  $\Omega\o\sigma$ &\cr
\tablerule
& B-type && $Z_D$   &&  $Z_C$   &\cr
 &  &&  $M$    &&  $W$    &\cr
\tablerule
& A-type && $Z_C$   &&  $Z_D$   &\cr
 &  &&  $W$    &&  $M$    &\cr
}\hrule}}}} 
\centerline{ \hbox{{\bf
Table 2:}{\it ~~ orientifold models}}} } 
\vskip 0.5cm
\noindent
Due to this relation via mirror symmetry we can restrict 
ourselves to the discussion of Type IIB orientifold models. 
Since in Type IIB one has even dimensional orientifold planes
and D-branes, one expects that the number of tadpole conditions
is related to the number of even cycles in the Calabi-Yau, i.e.
it is related to the Hodge number $h_{11}$. For the pure Gepner models  
with  $Z_D$ generically this number is rather small, whereas
for $Z_C$ it is rather big. Therefore, we expect that the A-type  
orientifold models are much more restrictive than the B-type  orientifolds.
It is known that by successive orbifolding one can reduce the number of 
$h_{11}$ while increasing $h_{21}$. Therefore, for such orbifold models
the A-type models might also become more tractable. 

For the B-type model it turns out that the resulting tadpoles
can be canceled by introducing B-type RS-boundary states, whereas
for the A-type model one uses A-type boundary states. 
After these general comments we are now in the position to
compute the Klein bottle amplitude for the B-type orientifold
models in detail. The result for the A-type Klein bottle
amplitude is briefly stated at the end of the next section.

\subsec{The B-type Klein bottle amplitude}

The general Klein bottle amplitude is defined as \foot{See for instance
\AngelantonjCT\ for a review on open string constructions.}
\eqn\klein{\eqalign{ K&=\int_0^\infty  {dt\over t} {\rm Tr}_{cl}\left(
             {\Omega\over 2} e^{-4\pi t \left(L_0-{c\over 24}\right)} \right) \cr 
          &={4\over c}\int_0^\infty  {dt\over t^3} {\rm Tr'}_{cl}\left(
             {\Omega\over 2} e^{-4\pi t \left(L_0-{c\over 24}\right)} \right) }}
where ${\rm Tr'}_{cl}$ denotes the trace over only the oscillator modes in the
closed string sector and
the integration over the
bosonic zero modes yields the factor $c=(8\pi^2 \alpha')^2$.
In the following we will set $c=1$.

Since all states from the range \range\ 
couple symmetrically in the diagonal Gepner partition function (B-type) 
\parti, they will all appear in the loop channel Klein bottle amplitude. 
Therefore, this amplitude can easily be written down
\eqn\kleina{  K^B={4}\int_0^\infty  {dt\over t^3}
         {1\over 2^{r+1}}  {1\over \eta(2it)^2 }
     {\sum_{\lambda,\mu}}^\beta (-1)^{s_0} \ \chi^\lambda_\mu (2it).  }
The next step is to transform this amplitude into the tree channel
by setting $t\to 1/(4l)$ and using a modular S-transformation. 
The computation is performed using the methods employed in 
\RecknagelSB\ for the computation of the annulus amplitudes. 
In order to carry out the summation over $\lambda$ and $\mu$
one has to extend  the sum ${\sum}^\beta$ over the entire
range \range\ by introducing $\delta$-functions written as
\eqn\deltaq{\eqalign{ \delta_{2\beta_0\bullet \mu\in 2\ZZ+1}&={1\over K}
    \sum_{\nu_0=0}^{K-1}  e^{i\pi\nu_0(q_{tot}-1)}, \cr
          \delta_{\beta_j\bullet \mu\in \ZZ}&=
         {1\over 2} \sum_{\nu_j=0}^{1}  e^{-i\pi\nu_j (s_0+s_j)}. }}           
After these steps we get
\eqn\kleinb{\eqalign{   \widetilde{K}^B=&\int_0^\infty  {dl}\, 
         {8\over 2^{3r}\,K }{1\over \prod_{j} (k_j+2)}
      {1\over \eta(2il)^2 }
     {\sum_{\lambda,\mu}}^{ev}{\sum_{\lambda',\mu'}}^{ev}\sum_{\nu_0=0}^{K-1}
     \sum_{\nu_1,\ldots,\nu_r=0}^1  (-1)^{\nu_0} \cr
     & e^{-i\pi {s_0\over 2} \left( s_0'+\nu_0 +2\sum \nu_j + 2\right)}\,\, 
     \prod_{j=1}^r \biggl( \sin (l_j,l'_j)_k\ 
     e^{i\pi {m_j\over k_j+2} \left( m_j'+\nu_0\right)}\,
     e^{-i\pi {s_j\over 2} \left( s_j'+\nu_0 +2 \nu_j \right)} \biggr)
   \,   \chi^{\lambda'}_{\mu'} (2il). }}
Taking into account that $l_j+m_j+s_j\in 2\ZZ$ we can carry out the sum
over $\lambda$ and $\mu$, which gives rise to some delta-functions.
After a few manipulations we can write the tree-channel Klein bottle amplitude
as 
\eqn\kleinc{\eqalign{   \widetilde{K}^B=\int_0^\infty  {dl}\, &
         {2^5\over 2^{r}\,K }
      {1\over \eta(2il)^2 }
     {\sum_{\lambda',\mu'}}^{ev}\sum_{\nu_0=0}^{K-1}
     \sum_{\nu_1,\ldots,\nu_r=0}^1  (-1)^{\nu_0}\, \delta_{s_0',2+\nu_0
       +2\sum \nu_j}^{(4)} \cr
    &\prod_{j=1}^r \biggl(  f(k_j,l'_j)\, 
    \delta_{m_j',\nu_0}^{(2k_j+4)}\, \delta_{s_j',\nu_0 +2 \nu_j}^{(4)} +
    g(k_j,l'_j)\, 
    \delta_{m_j',\nu_0+k_j+2}^{(2k_j+4)}\, \delta_{s_j',\nu_0 +2 \nu_j+2}^{(4)}
     \biggr)
   \,   \chi^{\lambda'}_{\mu'} (2il) }}
where we have introduced the notation
\eqn\ffgg{\eqalign{   f(k,l')&=\sum_{l=0}^{k} \sin (l,l')_k=
    \tan\left[{\pi\over 2}{(l'+1)(k+1)\over k+2}\right]\, \delta_{l',0}^{(2)}
   =\sqrt{k+2\over 2}\, {P_{l',k}^2\over S_{l',0}}, \cr
 g(k,l')&=\sum_{l=0}^{k} (-1)^{l} \sin (l,l')_k=f(k,k-l')=
      \sqrt{k+2\over 2}\, {P_{l',0}^2\over S_{l',0}}    .}}
The tree channel Klein bottle amplitude is also given by the
overlap of the so far unknown crosscap state $|C\ra_B$
\eqn\cross{  \widetilde{K}^B=\int_0^\infty  {dl} \la C|e^{-2\pi l H_{cl}} |
                C\ra_B .}
In order to extract this crosscap state (at least up to sign factors)
from the Klein bottle amplitude it is useful to rewrite 
\kleinc\ in terms of $S$ and $P$ matrices as 
\eqn\kleind{\eqalign{   \widetilde{K}^B={2^5  \prod_{j} \sqrt{k_j+2}
           \over 2^{{3r\over 2}} K } \,
         &\int_0^\infty  {dl}\, 
      {1\over \eta(2il)^2 }
     {\sum_{\lambda',\mu'}}^{ev} \sum_{\nu_0=0}^{K-1}
     \sum_{\nu_1,\ldots,\nu_r=0}^1  \sum_{\epsilon_1,\ldots,\epsilon_r=0}^1  
    (-1)^{\nu_0}\, \delta_{s_0',2+\nu_0
       +2\sum \nu_j}^{(4)} \cr
    &\prod_{j=1}^r \Biggl(  { P_{l'_j,\epsilon_j k_j}^2
     \over S_{l'_j,0} }  \, 
    \delta_{m_j',\nu_0+(1-\epsilon_j)(k_j+2)}^{(2k_j+4)}\, 
      \delta_{s_j',\nu_0 +2 \nu_j+2(1-\epsilon_j)}^{(4)}
     \Biggr)
   \,   \chi^{\lambda'}_{\mu'} (2il) }}
It can be checked that even though the sum is over the entire range \range,
the various $\delta$-functions in \kleind\ enforce that the 
GSO projection and the $\beta_j\bullet \mu\in\ZZ$ conditions
are  satisfied. From the final expression \kleind\ up to possible
signs, one can read off the form of the crosscap state
\eqn\crosscap{ \eqalign{ | C\ra_B={1\over \kappa_c}
   {\sum_{\lambda',\mu'}}^{ev} & \sum_{\nu_0=0}^{K-1}
     \sum_{\nu_1,\ldots,\nu_r=0}^1  \sum_{\epsilon_1,\ldots,\epsilon_r=0}^1  
     \Xi(\lambda',\mu',\nu_0,\nu_j,\epsilon_j)\, \delta_{s_0',2+\nu_0
       +2\sum \nu_j}^{(4)} \cr
    &\prod_{j=1}^r \Biggl(  {P_{l'_j,\epsilon_j k_j}
     \over \sqrt{S_{l'_j,0} }}  \, 
    \delta_{m_j',\nu_0+(1-\epsilon_j)(k_j+2)}^{(2k_j+4)}\, 
      \delta_{s_j',\nu_0 +2 \nu_j+2(1-\epsilon_j)}^{(4)}
     \Biggr)
   \,   |\lambda',\mu'\ra\ra_c. }}
 where we have suppressed the contribution from the two uncompactified
world-sheet bosons and where  
$|\lambda',\mu'\ra\ra_c$ denotes in the usual way the crosscap
Ishibashi states \CallanPX. 
For the overlap of two such Ishibashi states we choose in each tensor factor
\eqn\overl{  \la\la \tilde l,\tilde m,\tilde s | e^{-2\pi l (L_0+\o L_0-{c\over 12})}
        |l,m,s\ra\ra_c=\delta_{l,\tilde l}\,\delta_{m,\tilde m}\,\delta_{s,\tilde s}\,
            \chi^l_{m,s}(2il) .} 
From \kleind\ one can read off the normalization of the crosscap states
\eqn\nomiu{  {1\over \kappa_c^2}={2^5  \prod_{j=1}^r \sqrt{k_j+2}
           \over 2^{3r\over 2} K } .}
In section 5 we will determine the so far unknown sign factors
$\Xi(\lambda',\mu',\nu_0,\nu_j,\epsilon_j)$ from the consistency
of the M\"obius strip amplitude. 

A closer look at \kleind\ or \crosscap\ reveals
that only states are allowed to appear, which couple to their charge 
conjugate in the
diagonal torus partition function $Z_D$.
This is very reminiscent  to the B-type boundary states introduced
by Recknagel/Schomerus. This was the reason why we called
these orientifolds of the diagonal invariant B-type
orientifolds. 

\subsec{The A-type Klein bottle amplitude}

The computation for the charge conjugated modular invariant $Z_C$ (A-type)
is ana\-lo\-gous, so that we can keep its  presentation short.
In the loop channel only states which couple diagonally in 
$Z_C$ contribute to the trace. Apparently, these are the same states
which  couple to their charge conjugates in $Z_D$. 
As shown in \RecknagelSB, these states satisfy
\eqn\diag{    m_j=b\ {\rm mod}\ (k_j+2) }
for all $j$ and $b=0,\ldots,{K\over 2}-1$. 
Therefore, we can write the A-type Klein bottle amplitude as
\eqn\kleinaa{  K^A={4}\int_0^\infty  {dt\over t^3}
         {1\over 2^{r+1}}  {1\over \eta(2it)^2 }
     {\sum_{\lambda,\mu}}^\beta \sum_{b=0}^{{K\over 2}-1}
        \delta_{m_j,b}^{(k_j+2)}\, (-1)^{s_0} \ \chi^\lambda_\mu (2it).  }
Performing the same steps as for the B-type orientifold, one can transform 
this expression into the tree channel
\eqn\kleinda{\eqalign{   \widetilde{K}^A={2^4 
           \over 2^{{3r\over 2}} } \,&{1 \over \prod_{j} \sqrt{k_j+2}} 
         \int_0^\infty  {dl}\, 
      {1\over \eta(2il)^2 }
     {\sum_{\lambda',\mu'}}^{ev} \sum_{\nu_0=0}^{K-1}
     \sum_{\nu_1,\ldots,\nu_r=0}^1  \sum_{\epsilon_1,\ldots,\epsilon_r=0}^1  
    (-1)^{\nu_0}\, \delta_{s_0',2+\nu_0
       +2\sum \nu_j}^{(4)} \cr
   &\delta_{\sum_j {K'\over 2k_j+4}\left(m'_j+\nu_0+(1-\epsilon_j)(k_j+2)\right)}^{(K')}
    \prod_{j=1}^r \Biggl(  {P_{l'_j,\epsilon_j k_j}^2
     \over S_{l'_j,0} }  \, 
    \delta_{m_j'+\nu_0+(1-\epsilon_j)(k_j+2),0}^{(2)}\cr 
     & \delta_{s_j'+\nu_0 +2 \nu_j+2(1-\epsilon_j),0}^{(4)}
     \Biggr)
   \,   \chi^{\lambda'}_{\mu'} (2il) }}
Completely analogous to \crosscap\ this can be written as the overlap of a 
crosscap state
with normalization
\eqn\nomiua{  {1\over \kappa_c^2}={2^4 
           \over 2^{3r\over 2}  } {1\over \prod_{j=1}^r \sqrt{k_j+2}}.}
The $\delta$ functions in \kleinda\ do not pose any severe constraints, so
that essentially all states are allowed to contribute. Therefore, this 
A-type crosscap state couples to the same closed string modes as
the A-type RS-boundary states.

\newsec{RS boundary states}

Due to the general philosophy in orientifold model building, one now
has to introduce appropriate D-branes to cancel the R-R tadpoles
from the Klein bottle amplitude. 

\subsec{The B-type annulus amplitudes}

Since only states which couple to their
charge conjugates in $Z_D$ are present in the B-type tree channel
Klein bottle amplitude, it is clear that the suitable boundary states
to look at are the so-called B-type RS boundary states.
This means that these coherent states only contain Ishibashi states 
satisfying $q_i=-\o q_i$ and $h_i=\o h_i$ in each tensor factor.
Therefore, precisely those $\mu$ are allowed which satisfy
\eqn\bbb{   m_j=b\ {\rm mod}\ (k_j+2) }
for some $b=0,\ldots,{K\over 2}-1$ and all $j$. 
The complete B-type boundary state is given by
\eqn\btype{\eqalign{ |\alpha\ra_B=|S_0;(L_j,M_j,S_j)_{j=1}^r \ra_B=
         &{1\over \kappa^B_\alpha}
       {\sum_{\lambda',\mu'}}^{\beta,b} (-1)^{s_0^2\over 2}\,
          e^{-i\pi{s'_0\, S_0\over 2}} \cr
          &\prod_{j=1}^r {S_{l'_j,L_j}\over
       \sqrt{S_{l'_j,0} }}\, e^{i\pi{m'_j\, M_j\over k_j+2}}\,
      e^{-i\pi{s'_j\, S_j\over 2}} |\lambda',\mu'\ra\ra ,}}
where the sum is restricted to those GSO invariant states satisfying also
\bbb. It was shown in \RecknagelSB, that computing the overlap between two boundary states
of this type and transforming the resulting tree channel annulus amplitude
to loop channel via the S-transformation $l\to {1\over 2t}$ 
yields the following loop channel amplitude
\eqn\annul{\eqalign{   {A}^B_{\alpha\tilde\alpha}=
    &{N_{\alpha}\, N_{\tilde\alpha}\,
   \over 2^{{r\over 2}+1}\prod_j \sqrt{k_j+2}}{1\over  \kappa^B_{\alpha}
          \kappa^B_{\tilde\alpha} }
         \int_0^\infty  {dt\over t^3}\, 
      {1\over \eta(it)^2 }
     {\sum_{\lambda,\mu}}^{ev} \sum_{\nu_0=0}^{K-1}
     \sum_{\nu_1,\ldots,\nu_r=0}^1 \sum_{\epsilon_1,\ldots,\epsilon_r=0}^1    
    (-1)^{\nu_0}\cr
       & \delta_{s_0,2+\tilde S_0-S_0-\nu_0
       -2\sum \nu_j}^{(4)} \,\, 
         \delta_{
\sum_j {K'\over 2k_j+4}\left(m_j+M_j-\tilde M_j+\nu_0+\epsilon_j(k_j+2)\right)}^{(K')}
    \cr
    &\prod_{j=1}^r \Biggl( N_{L_j,\tilde L_j}^{|\epsilon_j k_j- l_j|}\, 
       \delta_{m_j+M_j-\tilde M_j+\nu_0+\epsilon_j(k_j+2),0}^{(2)}\,
       \delta_{s_j,\tilde S_j-S_j-\nu_0 -2 \nu_j+2\epsilon_j}^{(4)} \Biggr)
    \,   \chi^{\lambda}_{\mu} (it) }}
with $K'={\rm lcm}(k_j+2)$ and where we have also introduced the Chan-Paton factors $N_\alpha$.
The fusion matrix for the $SU(2)$ Kac-Moody algebra
is defined by the Verlinde formula
\eqn\fusion{   N_{i,j}^k=\sum_l  {S_{i,l}\, S_{j,l}\, S^*_{l,k}\over S_{0,l}}.}  
Note, that in contrast to \RecknagelSB\ we have written the annulus
amplitude in a manifest reflection symmetric way. 
The condition that in ${A}^B_{\alpha\alpha}$ the ground state appears with a factor
$1/2$  (due to the $(1+\Omega)/2$ projector) fixes the normalization factor to be
\eqn\normiu{   {1\over \left( \kappa^B_{\alpha}\right)^2}=
         {2\over 2^{r\over 2}K}\prod_{j=1}^r 
                  \sqrt{k_j+2} .}

\subsec{The A-type annulus amplitudes}

For the A-type boundary states only those Ishibashi states coupling 
in each tensor
to their charge conjugates in $Z_C$ are present.
These are the same states which couple diagonally in $Z_D$ satisfying 
$q_i=\o q_i$ and $h_i=\o h_i$ in each tensor factor. 
Therefore, the boundary state to be introduced to cancel the Klein bottle
tadpole are the so called A-type RS-boundary states
\eqn\atype{\eqalign{ |\alpha\ra_A=|S_0;(L_j,M_j,S_j)_{j=1}^r \ra_A=&{1\over \kappa^A_\alpha}
       {\sum_{\lambda',\mu'}}^{\beta} (-1)^{s_0^2\over 2}\,
          e^{-i\pi{s'_0\, S_0\over 2}} \cr
          &\prod_{j=1}^r {S_{l'_j,L_j}\over
       \sqrt{S_{l'_j,0}}}\, e^{i\pi{m'_j\, M_j\over k_j+2}}\,
      e^{-i\pi{s'_j\, S_j\over 2}} |\lambda',\mu'\ra\ra ,}}
which can be straightforwardly transformed into  loop channel 
\eqn\annula{\eqalign{   {A}^A_{\alpha\tilde\alpha}=N_{\alpha}\, N_{\tilde\alpha}\,
{\prod_j \sqrt{k_j+2}
    \over 2^{{r\over 2}}\, K} &{1\over  \kappa^A_{\alpha}
          \kappa^A_{\tilde\alpha} }
         \int_0^\infty  {dt\over t^3}\, 
      {1\over \eta(it)^2 }
     {\sum_{\lambda,\mu}}^{ev} \sum_{\nu_0=0}^{K-1}
     \sum_{\nu_1,\ldots,\nu_r=0}^1 \sum_{\epsilon_1,\ldots,\epsilon_r=0}^1    
    (-1)^{\nu_0}\cr
       & \delta_{s_0,2+\tilde S_0-S_0-\nu_0
       -2\sum \nu_j}^{(4)} \, 
    \prod_{j=1}^r \Biggl( N_{L_j,\tilde L_j}^{|\epsilon_j k_j- l_j|}\, 
       \delta_{m_j+M_j-\tilde M_j+\nu_0+\epsilon_j(k_j+2),0}^{(2k_j+4)}\cr
      & \delta_{s_j,\tilde S_j-S_j-\nu_0 -2 \nu_j+2\epsilon_j}^{(4)} \Biggr)
    \,   \chi^{\lambda}_{\mu} (it). }}
The normalization factor turns out to be 
\eqn\normiu{   {1\over \left( \kappa^A_{\alpha}\right)^2}=
         {K\over 2^{{r\over 2}+1}}{1\over \prod_{j=1}^r 
                  \sqrt{k_j+2}} .}
With all the annulus amplitudes available we would like to discuss whether in general
we can expect orientifolds of Gepner models to lead to  some
of the phenomenologically important issues of the Standard Model like
unitary gauge symmetries and chirality.

\subsec{Supersymmetry and chirality}

To begin with let us discuss what kinds of boundary states are generically 
allowed to be introduced into the orientifold background.
Requiring that the annulus amplitudes \annul\ and \annula\  only contain
NS-NS and R-R couplings between the different tensor factors implies 
$S_0-\tilde S_0$ and $S_j-\tilde S_j$ even.
Making the amplitude \annul\ self-consistent in the sense that only states with 
$l_j+m_j+s_j=0$ mod $2$  are allowed to contribute also requires $L_j+M_j+S_j=0$ mod $2$ 
for all $j$. Moreover, it is evident from the annulus amplitudes that
the boundary states also satisfy the reflection symmetry 
$(L_j,M_j,S_j)\to (k_j-L_j,M_j+k_j+2,S_j+2)$ in each tensor factor separately.
 
As we can see from the annulus amplitudes B-type branes
are classified by the combinations 
\eqn\combin{  M=\sum_j {K'\, M_j\over k_j+2}\ {\rm mod}\ 2K',\quad\quad S=\sum_j S_j .}
As was pointed put in \BrunnerJQ, the $S_j$ odd boundary states are sick in the
sense, that they do not yield consistent annulus amplitudes together with 
the $S_j$ even states. Moreover, as we will see in section 6 the crosscap state 
formally has $\tilde M_j=\tilde S_j=
\tilde S_0=0$ and therefore is an $S_j$ even state. 
Due to the reflection symmetry we can always choose $S_j=0$ for all $j$,
so that we are left with the two distinct possibilities $S_0\in\{0,2\}$.

Let us now determine what happens with a RS boundary state $|\alpha\ra_{A,B}$
under the world-sheet parity transformation $\Omega$. Since 
$|\alpha\ra_{B,A}$ contains Ishibashi states satisfying $q_i=-\o q_i$ from 
the $Z_{D,C}$ partition function, 
exchanging left and right movers leads to the following action
on a  boundary state
\eqn\action{ \Omega: |S_0;(L_j,M_j,S_j)_{j=1}^r \ra_{B,A} \to
            |-S_0;(L_j,-M_j,-S_j)_{j=1}^r \ra_{B,A} .}
Therefore, only states with $S_0=0,2$, $M_j=0$ and $S_j=0,2$ are actually invariant
under $\Omega$. All other states have to be introduced in pairs where
$\Omega$ maps a boundary state to the charge conjugated one. 
This means for the spectrum on the branes, that those 
which are invariant under $\Omega$ lead to $SO(N)$ and $SP(2N)$ gauge groups,
whereas all others give rise to $U(N)$ gauge groups. 
Thus we conclude, that unitary gauge groups are generically present in Gepner model
orientifolds. 

Now let us discuss the issue of supersymmetry. As in \RecknagelSB\ defining the charge 
\eqn\lad{ Q(\alpha)=-{S_0\over 2} -
         \sum_{j=1}^r {S_j\over 2} +\sum_{j=1}^r{M_j\over k_j+2}, }
the condition for two such boundary states to preserve the same supersymmetry reads
\eqn\susy{   Q(\alpha)-Q(\tilde\alpha)=0\ {\rm mod}\ 2 .}
Of course in the orientifold setting, in order to get a supersymmetric model
 the boundary states have also to preserve the same supersymmetry
as the crosscap state, which, as we have mentioned, 
formally has $\tilde M_j=\tilde S_j=
\tilde S_0=0$. 
In this paper we are only considering supersymmetric models, so that
we really require  $Q(\alpha)=0$ mod $2$  for all the D-branes we introduce. 

Then for B-type boundary states it is obvious that the supersymmetry
condition \susy\ with respect to the orientifold plane boils down
to $M=0$ mod $K'$. Therefore, all supersymmetric B-type boundary
states are invariant under the $\Omega$ projection and
carry always orthogonal or symplectic gauge groups.
Let us emphasize that for A-type boundary states such a condition does not arise
and they can also carry unitary gauge groups. 

In view of phenomenological applications, it is also important to know
whether the RS boundary states can lead to chiral models. 
The chiral matter content is geometrically given by the topological
(K-theoretic) intersection number between two boundary states. 
In the CFT this is given by the Witten index, which   has been
computed in \BrunnerJQ. For B-type boundary states the Witten index reads
\eqn\witten{   I_{\alpha,\tilde\alpha}=(-1)^{S-\tilde S\over 2 } \sum_{m_j'}
             \delta_{{M-\tilde M \over 2}+\sum {K'\over 2k_j+4}(m_j'+1),0}^{(K')}\,
             \prod_{j=1}^r N_{L_j,\tilde L_j}^{m'_j-1}  .}
Remember that for supersymmetric
D-branes we have $(M-\tilde M)=0$ mod $K'$. 
Now, inserting \susy\ in \witten\ one immediately realizes
that $I(\alpha,\tilde\alpha)=I(\tilde\alpha,\alpha)$. However,
one can generically prove that the intersection number 
is anti-symmetric, leaving only the possibility of vanishing 
intersection number between two relatively supersymmetric boundary states
\eqn\inter{   I_{\alpha,\tilde\alpha}=0 .}
Thus, we conclude that with just using the highly symmetric B-type RS-boundary
states one cannot built chiral supersymmetric orientifolds of Gepner 
models. This of course does not mean that on Calabi-Yau spaces, there are no
chiral intersecting D-brane models, it simply means that the 
set of RS-boundary states is too restrictive to achieve one
of the most salient features of the Standard Model.

For the A-type branes the Witten index is given by
\eqn\wittena{   I_{\alpha,\tilde\alpha}=(-1)^{S-\tilde S\over 2 } 
               \sum_{\nu_0=0}^{{K\over 2}-1}
             \prod_{j=1}^r N_{L_j,\tilde L_j}^{2\nu_0+M_j-\tilde M_j}  }
and in general it does not vanish for supersymmetric boundary states \foot{In 
\rbbkl\ it was found that for a subset of all supersymmetric A-type
branes in the $(3)^5$ Gepner model the Witten index actually vanishes,
but we have checked for instance in the $(1,1,7,7,7)$ Gepner model, 
that this is not a general rule.}. 

Summarizing, we have found that B-type orientifold models are less
constrained due to the smaller number of $h_{11}$, but that
their open string spectrum turns out to be phenomenologically
less appealing, as one always gets non-chiral models with
orthogonal or symplectic gauge groups. By turning
on the massless (anti-)symmetric matter on the branes one
can break the gauge symmetry down to unitary gauge factors, however at the
cost of  actually leaving the RS framework. Of course, even 
turning on these open string moduli does not lead to chirality.
Let us mention that these
models are very similar to the orientifolds with D-branes at angles
on toroidal orbifolds as discussed in \BlumenhagenMD.
In these latter models, the way to realize  unitary gauge groups and chirality
was to move to more general D-brane configurations, which led eventually
to the idea of intersecting brane worlds \rbgklnon. It would be interesting to
figure out whether more general boundary states can be defined
for Gepner models, maybe along the way proposed in \MizoguchiXI.

On the contrary, for the A-type models unitary gauge groups and chirality
are generic features of the open string spectrum. However, here the 
models become very complex, as there are a large number of
tadpole conditions to be satisfied by a very large set of possible
boundary states. Whether there really exist non-trivial chiral
models with large enough gauge group remains to be uncovered  and is
beyond the scope of this paper \timo. We end this section by stating
that these A-type models are very similar to the ordinary
orientifold models on toroidal orbifolds as first discussed in 
\refs{\GimonAY,\BerkoozDW}.

\newsec{The M\"obius strip amplitude}

The only  unoriented one-loop diagram which remains to be computed is
the M\"obius strip amplitude
\eqn\moebi{\eqalign{ M_\alpha&=\int_0^\infty  {dt\over t} {\rm Tr}_{\alpha\alpha'}\left(
             {\Omega\over 2} e^{-2\pi t \left(L_0-{c\over 24}\right)} \right) \cr 
          &={1\over c}\int_0^\infty  {dt\over t^3} {\rm Tr'}_{\alpha\alpha'}\left(
             {\Omega\over 2} e^{-2\pi t \left(L_0-{c\over 24}\right)} \right) }}
where the traces are over open strings stretching between the 
brane $\alpha$ and its $\Omega$ image $\alpha'$.
Generically, due to the $\Omega$ insertion one gets extra signs in the
M\"obius amplitude which can be described by the characters with 
shifted arguments  
$\chi(it+1/2)$. It is very convenient to introduce real characters defined as
\eqn\realm{  \hat\chi(it+1/2)=e^{-i\pi \left(h-{c\over 24}\right)}\, \chi(it+1/2) } 
and to express both the loop and tree channel M\"obius amplitude in terms of them. 
Formally the tree channel amplitude can be deduced from the boundary
and crosscap states as
\eqn\treemoeb{  \widetilde{M}_\alpha=\int_0^\infty  {dl} \la C|e^{-2\pi l H_{cl}} |
                \alpha\ra =\sum_j \Gamma_j\, B_j\,  \hat\chi_j(2il+1/2).}
where $\Gamma_j$ and $B_j$ are the crosscap respectively boundary state 
coefficients. 
In order to transform this to the loop channel one applies  the P-transformation
with $P=T^{1\over 2}ST^2ST^{1\over 2}$. 
Clearly in order to determine the tree channel M\"obius amplitude we have to 
know the complete crosscap state. In our case,  
so far we have only fixed the crosscap state up to  those extra signs 
$\Xi(\lambda',\mu',\nu_0,\nu_j,\epsilon_j)$ which we could not detect in the
Klein bottle amplitude. 
Our strategy to find them, is to impose the condition that GSO orbits of hatted
characters  transform into GSO orbits of hatted
characters  under the P-transformation. 

\subsec{P-transformation of orbits}

In this section we compute the P-transformation of a GSO orbit, where we are 
taking very carefully the extra signs into account. For our purposes it is sufficient
to just consider the NS sector amplitudes, which under the P-transformation transform
among  themselves and, as it turns out, are easier to handle in a general way. 
Assume that in some loop channel M\"obius amplitude 
the following GSO invariant orbit appears
\eqn\orbiti{  M^\lambda_\mu=\sum_{\nu=0}^{{K\over 2}-1} 
        \sum_{\nu_1,\ldots,\nu_r=0}^1 (-1)^{\left[ h^\lambda_{\mu}(\nu_0,\nu_j)-
         h^\lambda_{\mu}\right]}
    \,\, \widehat\chi^\lambda_{\mu+2\nu_0\beta_0+ \sum \nu_j \beta_j }
  (it+{\textstyle{1\over 2}}) ,}
where  $h^\lambda_{\mu}(\nu_0,\nu_j)$ denote the conformal dimensions
of the states appearing in the orbit and $h^\lambda_{\mu}=h^\lambda_{\mu}(0,\vec 0)$.
These extra signs appear in the amplitude after writing it in terms  of
 the hatted characters. 
Now we would like to figure  out what the resulting amplitude in the tree channel is by
applying  a P-transformation.  The appropriate P-matrix  \pmatrix\ introduces
a sign factor 
\eqn\signfact{   \prod_{j=1}^r \sigma_{(l,m,s)}=(-1)^{\left[
                   h^\lambda_{\mu}(\nu_0,\nu_j)-
         \Delta^\lambda_{\mu}(\nu_0,\nu_j)\right] } }
which combines with the sign in \orbiti\ in just the right way
to cancel the $(-1)^h$ factor and just leaves $(-1)^{\Delta}$.
The former sign is much harder to compute than the latter one, as it
requires appropriate reflections into the standard range \range.
However for the sign $(-1)^{\Delta}$ we can just use the general 
formula  \quannum\ and get the right result.
After some little algebra we find  
\eqn\delty{ \Delta^\lambda_{\mu}(\nu_0,\nu_j)-h^\lambda_{\mu}=
   \nu_0+\sum_{k<l} \nu_k \nu_l + \sum_j \nu_j\left( {s_0+s_j \over 2}+1 \right)\ 
        {\rm mod}\ 2 .}
The non-trivial quadratic piece $\sum_{k<l}   \nu_k \nu_l$ in \delty\ turns out 
to be quite important
for yielding indeed sums over orbits in  tree channel. 
In order to carry out the sums over the $\nu_j$ variables in \orbiti\ one has to evaluate
expressions like 
\eqn\nice{  F_r(\eta_1,\ldots,\eta_r)=
          \sum_{\nu_1,\ldots,\nu_r=0}^1 \left(\prod_{k<l}(-1)^{\nu_k \nu_l}\right)\,
                  e^{i\pi\left( \nu_1\eta_1+\ldots +\nu_r\eta_r\right)} ,}
which are slightly more complicated than what one is used to from the computation
of the Klein bottle and  annulus diagrams.
By performing the two sums over $\nu_r$ and $\nu_{r-1}$ explicitly, 
one derives the following recursion relation
\eqn\recur{  F_r(\eta_1,\ldots,\eta_r)=2\, (-1)^{\eta_{r-1}\eta_r}\, 
                   F_{r-2}(\eta_1+\eta_{r-1}+\eta_r+1,\ldots,\eta_{r-2}+\eta_{r-1}+\eta_r+1).}
Iteratively evaluating this relation, some inspection reveals that the
final answer for $F_r$ can be written as
\eqn\finalrel{\eqalign{  F_r(\eta_1,\ldots,\eta_r)=\cases{ (-1)^s\, 2^{r\over 2}\, 
                    \prod_{k<l} (-1)^{\eta_k \eta_l} \prod_{j} (-1)^{\eta_j} & for $r=4s$  \cr
                    (-1)^s\, 2^{r+1\over 2}\, 
                     \prod_{k<l} (-1)^{\eta_k \eta_l}\, \delta_{\sum_j \eta_j,0}^{(2)} 
              & for $r=4s+1$ \cr
                    (-1)^s\, 2^{r\over 2}\, 
                  \prod_{k<l} (-1)^{\eta_k \eta_l} & for $r=4s+2$  \cr
                  (-1)^s\, 2^{r+1\over 2}\, 
                   \prod_{k<l}  (-1)^{\eta_k \eta_l}\, 
                \delta_{\sum_j \eta_j,1}^{(2)} & for $r=4s+3$ \cr }}}
with $s\in \ZZ^+_0$. Since in our case $r=5$ or $r=9$, the second line in
\finalrel\ is relevant. 
Note, that the quadratic piece in \delty\ stays  form invariant under this 
discretized version of a Fourier transform. In this sense it is analogous
to the invariance of a Gaussian under a  continuous Fourier transformation.  
The form invariance of the quadratic piece
is very important for the tree and loop channel M\"obius amplitudes to contain the
correct relative signs in front of the different contributions in the GSO orbits.

Using \finalrel\ one finds after a few steps the following P-transformed amplitude
\eqn\treemoebi{\eqalign{   \widetilde{M}^\lambda_\mu\sim 
     {\sum_{\lambda',\mu'}}^{\beta}      
     \sum_{\epsilon_1,\ldots,\epsilon_r=0}^1  &
 \left( \prod_{j=1}^r \sigma_{(l_j',m_j',s_j')}\right)
     \, \left( \prod_{k<l} (-1)^{\eta_k \eta_l}\right)\,
                 \delta_{\sum_j \eta_j,0}^{(2)}\,\,
     e^{-i\pi{s_0\, s_0' \over 4}}\, \delta_{s_0+s_0',0}^{(2)}\cr
    &\prod_{j=1}^r \Biggl(  P_{l_j,|\epsilon_j k_j-l_j'|} 
       \,\, e^{i\pi{m_j\, m_j' \over 2k_j+4}}\,\,
      \delta_{m_j+m_j'+(1-\epsilon_j)(k_j+2),0}^{(2)}\,\,
          e^{-i\pi{s_j\, s_j' \over 4}}\,\,  \delta_{s_j+s_j',0}^{(2)}\cr
       &(-1)^{\epsilon_j {(m_j+s_j) \over 2}}\,
       (-1)^{\epsilon_j {(l'_j+m'_j+s'_j) \over 2}}\,
     \Biggr)
   \,   \widehat\chi^{\lambda'}_{\mu'} (2il+{\textstyle{1\over 2}}) }}
with 
\eqn\etatat{ \eta_j={s_0+s_j\over 2}-{s'_0+s'_j\over 2}+\epsilon_j+1.}
A little contemplation about this expression and checking it for some concrete 
examples reveals that 
indeed it contains again orbits over GSO invariant states with just the right
sign in front of each hatted character. 
After this little exercise, we can now come back to our orientifold models.  

\subsec{The B-type M\"obius amplitude}

Apparently the tree channel M\"obius strip amplitude \treemoebi\ has a 
very similar form than our tree channel Klein bottle amplitude.
Therefore, we choose \treemoebi\ as the guiding principle to fix the signs
in the crosscap state. Since we are eventually only considering supersymmetric models,
it is sufficient to  consider only the NS-part of the crosscap states.
Computing the $\eta_j$ in \etatat\ for
the restricted values of the Gepner model labels from \crosscap\
we find
\eqn\neww{    \eta_j=\sum_{i \ne j}  \nu_i .}
Now taking the signs from \treemoebi\ we finally get 
\eqn\crosscapb{ \eqalign{ | C\ra^{NS}_B={1\over \kappa_c}
   {\sum_{\lambda',\mu'}}^{ev}  \sum_{\nu_0=0}^{{K\over 2}-1}
     \sum_{\nu_1,\ldots,\nu_r=0}^1  \sum_{\epsilon_1,\ldots,\epsilon_r=0}^1  
     &(-1)^{\nu_0}\, \left(\prod_{k<l} (-1)^{\nu_k\nu_l} \right)\,
    (-1)^{\sum_j \nu_j}\, \delta_{s_0',2+2\nu_0
       +2\sum \nu_j}^{(4)} \cr 
    &\prod_{j=1}^r \Biggl(  \sigma_{(l_j',m_j',s_j')}\, 
      {P_{l'_j,\epsilon_j k_j}
     \over \sqrt{S_{l'_j,0} }}  \, 
    \delta_{m_j',2\nu_0+(1-\epsilon_j)(k_j+2)}^{(2k_j+4)}\cr 
     & \delta_{s_j',2\nu_0 +2 \nu_j+2(1-\epsilon_j)}^{(4)}\,
      (-1)^{\epsilon_j {(m'_j+s'_j) \over 2}}\, 
     \Biggr)
   \,   |\lambda',\mu'\ra\ra_c. }}
The next logical step is to compute the tree-channel M\"obius amplitude
between a boundary state and the crosscap state \crosscapb.
Since each Ishibashi state appearing in the crosscap state also appears
in the B-type boundary states, the tree-channel M\"obius amplitude turns
out to be
\eqn\moebiusb{ \eqalign{ \widetilde M^{B,NS}_{\alpha}=&-{2 N_\alpha\over \kappa_c\kappa_\alpha}
   \int_0^\infty dl\, {1\over \eta(2il+{1\over 2})^2 }
   {\sum_{\lambda',\mu'}}^{ev}  \sum_{\nu_0=0}^{{K\over 2}-1}
     \sum_{\nu_1,\ldots,\nu_r=0}^1  \sum_{\epsilon_1,\ldots,\epsilon_r=0}^1  
     (-1)^{\nu_0}\, \left(\prod_{k<l} (-1)^{\nu_k\nu_l} \right)\cr
    &(-1)^{\sum_j \nu_j}\,
    e^{i\pi {s'_0\, S_0\over 2}}\, \delta_{s_0',2+2\nu_0
       +2\sum \nu_j}^{(4)} 
   \prod_{j=1}^r \Biggl(  \sigma_{(l_j',m_j',s_j')}\, 
  {P_{l'_j,\epsilon_j k_j}
   \,  S_{l'_j,L_j}\over S_{l'_j,0} }  \, 
    \delta_{m_j',2\nu_0+(1-\epsilon_j)(k_j+2)}^{(2k_j+4)}\cr 
      &\delta_{s_j',2\nu_0 +2 \nu_j+2(1-\epsilon_j)}^{(4)}\,
      (-1)^{\epsilon_j {(m'_j+s'_j) \over 2}}\, e^{-i\pi {m'_j\, M_j\over k_j+2}}\,
      e^{i\pi {s'_j\, S_j\over 2}}\
     \Biggr)
   \,   \widehat\chi^{\lambda'}_{\mu'}(2il+{\textstyle{1\over 2}}), }}
where the overall sign has been fixed a posteriori by tadpole cancellation.  
Using the P-matrix with $l\to {1\over 8t}$  
this amplitude can be transformed  into loop channel.
After quite some algebra using for instance the formula
\eqn\lala{    Y^{k-l'}_{L,k-l}=(-1)^{2L+l-l'\over 2}\, Y^{l'}_{L,l}, }
with the $Y$ tensor defined below, 
we finally arrive at the following expression
\eqn\moebiloop{\eqalign{   {M}^{B,NS}_{\alpha}&=N_\alpha\,{(-1)^s\over 2^{r+1}}
         \int_0^\infty  {dt\over t^3}\, 
      {1\over \eta(it+{1\over 2})^2 }
     {\sum_{\lambda,\mu}}^{ev} 
     \sum_{\epsilon_1,\ldots,\epsilon_r=0}^1
     \left( \prod_{k<l} (-1)^{\rho_k\rho_l}\right)
       \delta_{\sum_j \rho_j,0}^{(2)}
    \, \delta_{s_0,0}^{(2)} \cr 
      &\delta_{\sum {K'\over 2k_j+4}\left[ 2M_j-m_j-
     \epsilon_j(k_j+2)\right],0}^{(K')} 
    \, \prod_{j=1}^r  \Biggl( \sigma_{(l_j,m_j,s_j)}\, \,
         Y_{L_j,\epsilon_j\, k_j}^{l_j}\, \,
       \delta_{m_j+\epsilon_j(k_j+2),0}^{(2)}\cr
      & \delta_{s_j,0}^{(2)}\, 
        (-1)^{{\epsilon_j\over 2}\left[2S_j-s_j-2\epsilon_j\right]}\,
        (-1)^{{(1-\epsilon_j)\over 2}
          \left[2M_j-m_j-\epsilon_j(k_j+2)\right]}\Biggr)
    \,   \widehat\chi^{\lambda}_{\mu} (it+{\textstyle{1\over 2}}) }}
with $r=4s+1$ and 
\eqn\rhoo{  \rho_j={s_0+s_j\over 2} + \epsilon_j-1.}
In \moebiloop\ the integer valued $Y_{l_1,l_2}^{l_3}$ tensor  of $SU(2)_k$ is defined
as 
\eqn\ymatr{  Y_{l_1,l_2}^{l_3}=\sum_{l=0}^k  {S_{l_1,l}\, P_{l_2,l} \, P_{l_3,l}\over
                               S_{0,l} } .}   
Note, that the sign factor $(-1)^s$ gives rise to different projections depending
on the number of tensor factors. 
For $r=5$ one always gets orthogonal  gauge groups, whereas for  the $(1)^9$ 
Gepner model one finds symplectic gauge groups. 
Evaluating  the general formula for the M\"obius amplitude we can now revisit
from a formal point of view the question about 
the kinds of gauge groups living on the RS-boundary states.
Recall    that relatively supersymmetric D-branes must satisfy
$M=0$ mod $K'$. Inserting this into \moebiloop\ one realizes from the
$\delta^{(K')}$ constraint that the ground state $(2)(0,0,0)^5$ always
appears. Therefore, we can confirm our  general expectation that for all supersymmetric  
D-branes the
gauge sector gets symmetrized or anti-symmetrized  and  no
unitary gauge groups are possible.

\subsec{The A-type M\"obius amplitude}

For the A-type crosscap and boundary states the overlap is
\eqn\moebiusa{ \eqalign{ \widetilde M^{A,NS}_{\alpha}=&-{2N_\alpha\over \kappa^A_c\kappa^A_\alpha}
   \int_0^\infty dl\, {1\over \eta(2il+{1\over 2})^2 }
   {\sum_{\lambda',\mu'}}^{ev}  \sum_{\nu_0=0}^{{K\over 2}-1}
     \sum_{\nu_1,\ldots,\nu_r=0}^1  \sum_{\epsilon_1,\ldots,\epsilon_r=0}^1  
     (-1)^{\nu_0}\, \left(\prod_{k<l} (-1)^{\nu_k\nu_l} \right)\cr
    &(-1)^{\sum_j \nu_j}\,
    e^{i\pi {s'_0\, S_0\over 2}}\, \delta_{s_0',2+2\nu_0
       +2\sum \nu_j}^{(4)} \,\,
  \delta_{\sum_j {K'\over 2k_j+4}\left(m'_j+2\nu_0+(1-\epsilon_j)(k_j+2)\right)}^{(K')}\cr
  &\prod_{j=1}^r \Biggl(  \sigma_{(l_j',m_j',s_j')}\, 
 {P_{l'_j,\epsilon_j k_j}
   \,  S_{l'_j,L_j}\over S_{l'_j,0} }  \, 
    \delta_{m_j',2\nu_0+(1-\epsilon_j)(k_j+2)}^{(2)}\cr 
      &\delta_{s_j',2\nu_0 +2 \nu_j+2(1-\epsilon_j)}^{(4)}\,
      (-1)^{\epsilon_j {(m'_j+s'_j) \over 2}}\, e^{-i\pi {m'_j\, M_j\over k_j+2}}\,
      e^{i\pi {s'_j\, S_j\over 2}}\
     \Biggr)
   \,   \widehat\chi^{\lambda'}_{\mu'}(2il+{\textstyle{1\over 2}}). }}
The transformation into loop channel yields
\eqn\moebiloopa{\eqalign{   {M}^{A,NS}_{\alpha}&=N_\alpha\,{(-1)^s\over 2^{r+1}}
         \int_0^\infty  {dt\over t^3}\, 
      {1\over \eta(it+{1\over 2})^2 }
     {\sum_{\lambda,\mu}}^{ev} 
     \sum_{\epsilon_1,\ldots,\epsilon_r=0}^1
     \left( \prod_{k<l} (-1)^{\rho_k\rho_l}\right)
       \delta_{\sum_j \rho_j,0}^{(2)}
    \, \delta_{s_0,0}^{(2)} \cr 
      & \prod_{j=1}^r  \Biggl( \sigma_{(l_j,m_j,s_j)}\, \,
         Y_{L_j,\epsilon_j\, k_j}^{l_j}\, \,
       \delta_{2M_j-m_j-\epsilon_j(k_j+2),0}^{(2k_j+4)}\,
       \delta_{s_j,0}^{(2)}\, 
        (-1)^{{\epsilon_j\over 2}\left[2S_j-s_j-2\epsilon_j\right]}\cr
       & (-1)^{{(1-\epsilon_j)\over 2}
          \left[2M_j-m_j-\epsilon_j(k_j+2)\right]}  \Biggr)
    \,   \widehat\chi^{\lambda}_{\mu} (it+{\textstyle{1\over 2}}). }}
Having all one-loop amplitudes available we can move forward and  
compute the tadpoles and from the loop channel amplitudes the
massless spectra.  Note that in contrast to earlier approaches to the
construction of orientifolds of Gepner models, we have derived
absolutely general formulas for all the relevant one-loop amplitudes,
which can be easily installed on a computer.

\newsec{Tadpole cancellation}

Massless states propagating between the various combinations
of crosscap and boundary states
give rise to divergences in the respective one-loop diagram. 
For a consistent string model  one has to require that after adding
up all 1-loop diagrams these divergences do cancel.

\subsec{B-type models}

For the B-type models only those states $(\lambda,\mu)$ are present  in the crosscap and
boundary states, which appear in $Z_D$ in combination with their
charge conjugate ones $(\lambda,-\mu)$. Therefore, besides the
vacuum only some of the chiral-antichiral, $(ac)$, states can lead  to tadpoles. 
This implies that the number of independent tadpole cancellation conditions
is at most $N_{ac}+1=h_{11}+1$. 

In order to cancel the tadpoles from the orientifold planes, we are introducing 
stacks of B-type boundary states
$|S_0;\prod_j (L^a_j,0,0)\ra$ with $L_j$ even and CP-factors $N_a$.
By extracting the divergences from the
various one-loop amplitudes, we find that, as required, they lead to perfect squares.
The contribution from a massless state $(\lambda,\mu)$ is given by
the intriguingly simple expression
\eqn\tadpole{ {2\over K}{2^{r}\over \prod_j \sin (l_j,0)_{k_j}} 
         \left( \sum_{a=1}^N N_a \prod_j \sin (l_j,L^a_j)_{k_j} - 4 
                \prod_j \sin {1\over 2}(l_j,k_j)_{k_j} \right)^2 =0.}
Using the field identifications, 
without loss of generality we have assumed that all $m_j$ are even.
Moreover, $N$ denotes the number of stacks of different boundary states and
$N_a$ the number on each individual stack.

From the general tadpole cancellation conditions \tadpole\ it is immediately
clear that there always exists a simple solution to these equations namely
by choosing one stack of D-branes with
\eqn\simplesol{    L_j={k_j\mp1\over 2}  }
for all $j$ and $k_j=4n_j\pm 1$.
The Chan-Paton factor
is just $N_1=4$ and for $r=5$  leads to a gauge group $SO(4)$
and for $r=9$ to $SP(4)$. 
The interpretation of this solution is that we have just placed 
appropriate D-branes right on top of the
orientifold plane.
From the phenomenological point of view this simple
solution is not very interesting for its rank is far to small 
to accommodate the Standard Model gauge symmetry. 
As we will show in the next section, the set of equations
\tadpole\ allows more general solutions, which can also have 
amazingly high rank gauge symmetries.

\subsec{A-type models}

For A-type models massless tadpoles can arise from states
which couple to their charge conjugates in $Z_C$ respectively 
diagonally in $Z_D$. Therefore the number of tadpole
cancellation conditions is given by $h_{21}$ for the  $Z_D$ model.
As in the B-type model we introduce A-type boundary states
$|S_0;\prod_j (L^a_j,M_j,S_j)\ra$ with CP-factors $N_a$.
As we have argued, here also non-zero values
of $M_j$ and $S_j$ are allowed so that in general we have to introduce
also the $\Omega$ image brane $|S_0;\prod_j (L^a_j,-M_j,-S_j)\ra$.
By shifting again the labels such that all $m_j$ are even we
obtain for the A-type tadpole cancellation the general
formula
\eqn\tadpolea{\eqalign{  
         \Biggl( \sum_{a=1}^N 2\, N_a\, \cos\biggl[\pi 
         {\textstyle{s_0\, S^a_0\over 2}}-\pi\sum_j\left( 
    {\textstyle{m_j\, M^a_j\over k_j+2}}
              +{\textstyle{s_j\, S^a_j\over 2}}\right) \biggr]\,
          &\prod_j \sin (l_j,L^a_j)_{k_j}  \cr
                  &   -4 
                \prod_j \sin {1\over 2}(l_j,k_j)_{k_j} \Biggr)^2 =0.}}
For $\Omega$ invariant boundary states the $\cos [\ ]$ term vanishes and
the overall CP-factor is $2N_a$. 
From \tadpolea\ it is clear that there also exists a generic solution 
corresponding to placing 
four branes just on top of the orientifold plane
\eqn\simplesola{    L_j={k_j\mp1\over 2},\ M_j=S_j=S_0=0  }
for all $j$ and $k_j=4n_j\pm 1$.
Here again we get $SO(4)$ gauge group for $r=5$ and $SP(4)$ for $r=9$.
It would be interesting to evaluate these tadpole cancellation
conditions for non $\Omega$ invariant boundary states and see whether
non-trivial solutions exist \timo.

\newsec{Examples}

In this section we discuss a couple of B-type examples in more detail. 
We focus in this paper on B-type examples, where quite easily
non-trivial solutions can be found. For the A-type models
we have not yet managed to find any non-trivial solutions
to the tadpole cancellation conditions and leave a more
thorough analysis for future work.
Since it is the easiest non-trivial example let us revisit first the $(3)^5$ Gepner model.

\subsec{Orientifolds of $(3)^5$ Gepner model}

In order to evaluate  the general tadpole conditions \tadpole\ we first have
to determine what kinds of massless tadpoles can occur. For the $(3)^5$ 
Gepner model  only two massless $(ac)$ states $(s_0)\prod_j (l_j,m_j,s_j)$ 
and their charge conjugates do exist, namely 
\eqn\massless{\eqalign{   (2)(0,0,0)^5\,\quad {\rm and}\  (0)(1,1,0)^5=(0)(2,-4,2)^5 ,}}
where for the second state we have applied the reflection symmetry to
guarantee $m_j$ even. 
The next step is to decide what kinds of RS-boundary states one wants 
to introduce in order to cancel the two tadpoles. 
In principle one is free to introduce any set of B-type boundary states
and try to cancel the tadpoles with them. 
We choose here  the set of B-type boundary states with $S_0=0$ displayed
in Table 3.
\vskip 0.8cm
\vbox{ \centerline{\vbox{ \hbox{\vbox{\offinterlineskip
\def\tablespace{height2pt&\omit&&\omit&&\omit&&
 \omit&\cr}
\def\tablerule{\tablespace\noalign{\hrule}\tablespace}

\hrule\halign{&\vrule#&\strut\hskip0.2cm\hfill #\hfill\hskip0.2cm\cr
& RS-boundary state &&    &&   && &\cr
\tablespace
& $(L_1,L_2,L_3,L_4,L_5)$ &&  degeneracy  &&  CP-factor && moduli $n_{(S,A)}$ &\cr
\tablerule
& $(0,0,0,0,0)$ && $1$ && $N_0$ && $0$ &\cr
\tablerule
& $(0,0,0,0,2)$ && $5$ && $N_1$ && $4_{(4,0)}$ &\cr
\tablerule
& $(0,0,0,2,2)$ && $10$ && $N_2$ && $11_{(8,3)}$ &\cr
\tablerule
& $(0,0,2,2,2)$ && $10$ && $N_3$ && $24_{(15,9)}$ &\cr
\tablerule
& $(0,2,2,2,2)$ && $5$ && $N_4$ && $50_{(28,22)}$ &\cr
\tablerule
& $(2,2,2,2,2)$ && $1$ && $N_5$ && $101_{(51,50)}$ &\cr
}\hrule}}}} 
\centerline{ \hbox{{\bf
Table 3:}{\it ~~ boundary states}}} } 
\vskip 0.5cm
\noindent
In Table 3 we have also given the dimension of the open string moduli
space of each boundary state and, utilizing the loop channel M\"obius 
amplitude, we have also indicated  how these scalars transform under
the $SO(N_j)$ gauge symmetry. 
Inserting these values in the tadpole cancellation conditions
\tadpole, one realizes that they can be written in a compact form as
\eqn\quintad{\eqalign{ &\sum_{i=0}^5 N_i \kappa^i=4 \kappa^5 \cr
                       &\sum_{i=0}^5 (-1)^i\, N_i \kappa^{-i}=-4 \kappa^{-5} }} 
with $\kappa={1+\sqrt{5}\over 2}$. Adding these two equations one gets
\eqn\addia{ 2N_0 + N_1 + 3 N_2 + 4 N_3 + 7 N_4 + 11 N_5=44 .}
Eliminating $N_5$ from \quintad\ yields a second equation
\eqn\addib{ 5N_0 - 3 N_1 + 2 N_2 -  N_3 +  N_4 =0 .}
As expected $N_5=4$ with all other CP-factors vanishing is a solution
to these two equations. However, there are many others with even larger
rank of the gauge group. The largest rank we can get is sixteen 
and is given by the choice $N_0=12$, $N_1=20$ which leads to a gauge group
\eqn\gaut{  G=SO(12)\times SO(20) .}
This result has first been derived for a freely acting  orbifold of the 
$(3)^5$ Gepner model 
in \BlumenhagenTJ\  and has later been confirmed for the $(3)^5$ model itself in 
\AldazabalUB. It is quite remarkable that such high rank solutions do exist and we will
confirm their existence also for the other Gepner models discussed in this
section.

Given a solution to the tadpole equations one can evaluate the
loop channel annulus and M\"obius strip amplitudes and read off the massless
spectrum. For the model with maximal gauge symmetry above, one finds additional
matter multiplets in the following representations
\eqn\matter{  4\times ({\bf 1},{\bf 210})+4\times ({\bf 12},{\bf 20}) .}
Note, that the boundary state $(0,0,0,0,0)$ is rigid in the sense that
it does not have any open string moduli space.  This  property  has turned out 
to be  really hard to realize  in toroidal orbifold models. From the phenomenological 
point of view rigid cycles are very attractive, as they do not
lead to additional adjoint (anti-symmetric) matter, which might
spoil the nice gauge unification properties of the MSSM \BlumenhagenJY.

\subsec{Orientifolds of $(1)^2\, (7)^3$ Gepner model}

As a second example we discuss the $(1)^2\, (7)^3$ Gepner model.
Since $h_{11}=4$ we expect at most five tadpole cancellation conditions.
However, a deeper look into the $(ac)$ states of the model reveals
that only three of the possible five states really appear in the
B-type boundary and crosscap states 
\eqn\massless{\eqalign{  &(2)(0,0,0)^5 \cr
                         & (0)(1,1,0)^5=(0)(0,-2,2)^2\, (6,-8,2)^3 \cr
                         &(0)(0,0,0)^2\, (3,3,0)^3=(0)(0,0,0)^2(4,-6,2)^3 .}}
The other two $(ac)$ states are left-right combinations of the form
\eqn\acstate{  (0)(1,1,0)(0,0,0)(2,2,0)^3_L\bowtie (0)(0,0,0)(1,-1,0)(2,-2,0)^3_R }
and as we said do not couple to the boundary and crosscap states.
Now we are introducing the boundary states listed in Table 4.
It is quite impressive that the initial three tadpole cancellation conditions
with in general irrational coefficients can be linear transformed, so that
finally one ends up with three equations with only integer  coefficients.
After some algebra we can write the tadpole cancellation conditions as
\eqn\tadzwei{\eqalign{  32=& N_0 +  N_1 +  N_2 + 3 N_4 + 2 N_5 + N_6 +
              3 N_7 + 2 N_8 + 2 N_9 + 6 N_{10} + 6 N_{11} + \cr
          &3 N_{12} +  7 N_{13} + 5 N_{14} + 4 N_{15} + 8 N_{16}  + 5 N_{17}+
               3 N_{18} +  N_{19},  \cr
              0=&2 N_1 -  N_2 -N_3 +  N_4 -2  N_6 +
              N_8 + 2 N_9 + 3 N_{10} -  N_{11} -2  N_{12}
              + N_{13} +\cr
                   & N_{14} + 3 N_{15} 
               - N_{18} - 3 N_{19}, \cr
              0=&-6 N_0 - N_2 + 5 N_3 -7  N_4 + 4 N_5 +4 N_6 
              -2  N_7 -  N_8 - 6 N_9 -3  N_{10} +  N_{11} +\cr 
            &8 N_{12} +
              N_{13} -3  N_{14} -7  N_{15} + 2 N_{17}+
               3 N_{18} + 9 N_{19}  .}}
The trivial solution is here $N_{16}=4$, but there again exist solutions
with larger rank. Choosing for instance $N_0=8$ and $N_5=12$ with all
other CP-factors vanishing yields a gauge group 
\eqn\gauget{   G=SO(8)\times SO(12) }
of rank rk$(G)=10$.
Evaluating the loop channel annulus and M\"obius amplitudes, one finds additional
matter multiplets in the following representations
\eqn\matter{  13\times ({\bf 1},{\bf 78})+7\times ({\bf 1},{\bf 66})+
3\times ({\bf 8},{\bf 12}) .}

A second solution with even larger rank is given by the choice 
$N_0=8$, $N_3=24$ and $N_9=12$ leading  to a gauge group of rank rk$(G)=22$(!)
\eqn\gaugetb{   G=SO(8)\times SO(24) \times SO(12) .}
The resulting massless matter transforms in the following representations
\eqn\matterb{ 
2\times ({\bf 1},{\bf 300},{\bf 1})+4\times ({\bf 1},{\bf 1},{\bf 78})+
2\times ({\bf 8},{\bf 24},{\bf 1 })+1\times ({\bf 8},{\bf 1},{\bf 12 })+
n_{39} \times ({\bf 1},{\bf 24},{\bf 12 }) }
with $n_{39}=5$ for $\vec L_3=(0,0,6,0,0)$ and   $\vec L_9=(0,0,6,6,0)$
and $n_{39}=1$ for $\vec L_3=(0,0,6,0,0)$ and   $\vec L_9=(0,0,0,6,6)$.

\vskip 0.8cm
\vbox{ \centerline{\vbox{ \hbox{\vbox{\offinterlineskip
\def\tablespace{height2pt&\omit&&\omit&&\omit&&
 \omit&\cr}
\def\tablerule{\tablespace\noalign{\hrule}\tablespace}

\hrule\halign{&\vrule#&\strut\hskip0.2cm\hfill #\hfill\hskip0.2cm\cr
& RS-boundary state &&    &&  &&  &\cr
\tablespace
& $(L_1,L_2,L_3,L_4,L_5)$ &&  degeneracy  &&  CP-factor && moduli $n_{(S,A)}$ &\cr
\tablerule
& $(0,0,0,0,0)$ && $1$ && $N_0$ && $0$ &\cr
\tablerule
& $(0, 0, 2, 0, 0)$ && $3$ && $N_1$ && $3_{(2,1)}$ &\cr
\tablerule
& $(0, 0, 4, 0, 0)$ && $3$ && $N_2$ && $5_{(4,1)}$ &\cr
\tablerule
& $(0, 0, 6, 0, 0)$ && $3$ && $N_3$ && $2_{(2,0)}$ &\cr
\tablerule
& $(0, 0, 2, 2, 0)$ && $3$ && $N_4$ && $14_{(10,4)}$ &\cr
\tablerule
& $(0, 0, 4, 2, 0)$ && $6$ && $N_5$ && $20_{(13,7)}$ &\cr
\tablerule
& $(0, 0, 6, 2, 0)$ && $6$ && $N_6$ && $8_{(7,1)}$ &\cr
\tablerule
& $(0, 0, 4, 4, 0)$ && $3$ && $N_7$ && $27_{(16,11)}$ &\cr
\tablerule
& $(0, 0, 6, 4, 0)$ && $6$ && $N_8$ && $13_{(9,4)}$ &\cr
\tablerule
& $(0, 0, 6, 6, 0)$ && $3$ && $N_9$ && $4_{(4,0)}$ &\cr
\tablerule
& $(0, 0, 2, 2, 2)$ && $1$ && $N_{10}$ && $45_{(35,10)}$ &\cr
\tablerule
& $(0, 0, 4, 2, 2)$ && $3$ && $N_{11}$ && $62_{(40,22)}$ &\cr
\tablerule
& $(0, 0, 6, 2, 2)$ && $3$ && $N_{12}$ && $29_{(25,4)}$ &\cr
\tablerule
& $(0, 0, 4, 4, 2)$ && $3$ && $N_{13}$ && $84_{(47,37)}$ &\cr
\tablerule
& $(0, 0, 6, 4, 2)$ && $6$ && $N_{14}$ && $41_{(28,13)}$ &\cr
\tablerule
& $(0, 0, 6, 6, 2)$ && $3$ && $N_{15}$ && $18_{(17,1)}$ &\cr
\tablerule
& $(0, 0, 4, 4, 4)$ && $1$ && $N_{16}$ && $112_{(57,55)}$ &\cr
\tablerule
& $(0, 0, 6, 4, 4)$ && $3$ && $N_{17}$ && $56_{(33,23)}$ &\cr
\tablerule
& $(0, 0, 6, 6, 4)$ && $3$ && $N_{18}$ && $26_{(19,7)}$ &\cr
\tablerule
& $(0, 0, 6, 6, 6)$ && $1$ && $N_{19}$ && $11_{(11,0)}$ &\cr
}\hrule}}}} 
\centerline{ \hbox{{\bf
Table 4:}{\it ~~ boundary states for (1,1,7,7,7) Gepner model}}} } 
\vskip 0.5cm
\noindent

\subsec{Orientifold of $(1)\, (3)^3\, (13)$ Gepner model}

The last example we want to discuss is the $(1)\, (3)^3\, (13)$ Gepner model
which has $h_{11}=3$ so that we expect at most four tadpole constraints.
Indeed the $(ac)$ states which are coupling to the boundary and
crosscap states are
\eqn\massless{\eqalign{   &(2)(0,0,0)^5 \cr
                          &(0)(1,1,0)^5=(0,-2,2)(2,-4,2)^3 (12,-14,2)\cr
                          &(0)(0,0,0)(1,1,0)^3(6,6,0)=(0)(0,0,0)(2,-4,2)^3(6,6,0)\cr
                          &(0)(1,1,0)(0,0,0)^3(10,10,0)=(0)(0,-2,2)(0,0,0)^3(10,10,0) .}}
For this example we are here only considering the boundary states listed
in Table 5. A more complete list of boundary states and the resulting
tadpole cancellation conditions can be found in Appendix A.  
\vskip 0.8cm
\vbox{ \centerline{\vbox{ \hbox{\vbox{\offinterlineskip
\def\tablespace{height2pt&\omit&&\omit&&\omit&&
 \omit&\cr}
\def\tablerule{\tablespace\noalign{\hrule}\tablespace}

\hrule\halign{&\vrule#&\strut\hskip0.2cm\hfill #\hfill\hskip0.2cm\cr
& RS-boundary state &&    &&   && &\cr
\tablespace
& $(L_1,L_2,L_3,L_4,L_5)$ &&  degeneracy  &&  CP-factor && moduli $n_{(S,A)}$ &\cr
\tablerule
& $(0,0,0,0,0)$ && $1$ && $N_0$ && $0$ &\cr
\tablerule
& $(0, 0, 0, 0, 2)$ && $1$ && $N_1$ && $0$ &\cr
\tablerule
& $(0, 2, 0, 0, 4)$ && $3$ && $N_2$ && $8_{(7,1)}$ &\cr
\tablerule
& $(0, 0, 0, 0, 6)$ && $1$ && $N_3$ && $7_{(4,3)}$ &\cr
\tablerule
& $(0, 2, 0, 0, 6)$ && $3$ && $N_4$ && $15_{(9,6)}$ &\cr
\tablerule
& $(0, 2, 2, 0, 6)$ && $3$ && $N_5$ && $31_{(17,14)}$ &\cr
\tablerule
& $(0, 2, 2, 2, 6)$ && $1$ && $N_6$ && $63_{(32,31)}$ &\cr
\tablerule
& $(0, 2, 0, 0, 10)$ && $3$ && $N_7$ && $8_{(7,1)}$ &\cr
\tablerule
& $(0, 0, 0, 0, 12)$ && $1$ && $N_8$ && $0$ &\cr
}\hrule}}}} 
\centerline{ \hbox{{\bf
Table 5:}{\it ~~ boundary states for the (1,3,3,3,13) Gepner model}}} } 
\vskip 0.5cm
\noindent
For this choice of boundary states, it turns out that only three out of 
the four tadpole conditions are actually independent 
\eqn\taddrei{\eqalign{  72=& 4 N_0 +  5 N_1 +  3 N_2 + 4 N_3 + 7 N_4 + 11 N_5 + 18 N_6 +
              8  N_7 - N_8  \cr
              0=& N_0 + N_1 -  N_2 + N_7 - N_8 \cr
              0=&5 N_1 - N_2 + 2 N_3 -  N_4 +  N_5 +2 N_7 
              -3  N_8. \cr  }}
Choosing for instance
\eqn\wahnsinn{  N_0=8, N_1=12, N_8=20 }
with the remaining CP-factors vanishing leads to the gauge group
\eqn\gauget{   G=SO(8)\times SO(12)\times SO(20) }
of rank rk$(G)=20$. 
For this model one finds additional
matter multiplets in the following representations
\eqn\matter{  3\times ({\bf 8},{\bf 1},{\bf 20})+4\times ({\bf 1},{\bf 12},{\bf 20}) .}

We hope that these examples have convinced the reader that 
 orientifolds of Gepner models though complicated can still be constructed in a systematic way
and that they admit  surprising and new patterns of solutions to the tadpole
conditions.

\newsec{Conclusions}

In this paper we have developed a systematic approach for constructing
orientifolds of Gepner models. Extending previous work,
at least for odd levels, we presented quite general expressions for all
A-type and B-type one-loop amplitudes.  From these amplitudes 
we derived the general
form of the tadpole cancellation conditions, which in the end 
turned out to be fairly simple. 
Finally, we have also discussed a couple of B-type examples
in some more detail, in particular confirming that solutions
of higher rank generically exist. Moreover, we pointed out on general
grounds that
for relatively supersymmetric B-type boundary states only orthogonal
or symplectic gauge groups with necessarily non-chiral massless matter
are possible.

This work is meant to be just a first step into an even more
general investigation of orientifolds of Gepner models, where
one would also include the cases with even levels and just four
tensor factors. 
The hope is, that for these more general  cases some of the
limitations of the B-type type models
might be overcome. 
It might also be possible that one has to move beyond the
RS-boundary states to realize these phenomenologically appealing
features in B-type orientifolds of Gepner models. 

Moreover, it remains to be seen whether non-trivial examples
of A-type orientifold models do exist, where one can indeed get unitary
gauge groups and chiral fermions. One possibility in order to reduce the number
of conditions is to consider  additional orbifolds or simple current
constructions \refs{\BlumenhagenTJ,\AldazabalUB}. 
Encouragingly,  for a simple current extension
of the A-type orientifold of $(1,1,7,7,7)$ Gepner model one example of a 
higher rank gauge group was found in \BlumenhagenTJ. 
From the phenomenological point of view, the final aim of all these
efforts would be to
systematically analyze such orientifold models for their 
ability to yield models which come close to the Standard Model. 
After all, so far we have just revealed the tip of the
iceberg of the whole plethora of Gepner model orientifolds. 

It would also be interesting to see whether the orientifold
models constructed in this paper are dual after compactification on a circle
to the SCFTs for $G_2$ manifolds
proposed in \BlumenhagenJB.

\vskip 1cm
\centerline{{\bf Acknowledgements}}\pano
This work  is supported by PPARC. I would like to thank A. Wi\ss kirchen
for helpful  correspondence and T. Weigand for useful comments about the
manuscript. 

\vfill\eject
\appendix{A}{Tadpole conditions for $(1,3,3,3,13)$ Gepner model}

In this appendix we present the general set of boundary states for the
B-type orientifold of the $(1,3,3,3,13)$ Gepner model. 
Introducing the boundary states listed in Table 6
the four tadpole cancellation conditions read 
\eqn\appa{\eqalign{ 72 = &4 N_0+2 N_1 +6 N_2 + 8 N_3 + 
  5 N_4 + 5 N_5 + 10 N_6 + 15 N_7 + 6 N_8 + 3 N_9 +
   9 N_{10} + 12 N_{11} + \cr
   & 4 N_{12} + 7 N_{13}+ 
  11 N_{14} + 18 N_{15} + 5 N_{16} + 
  10 N_{17} + 15 N_{18} + 25 N_{19}+
   N_{20} + 8 N_{21} + \cr & 9 N_{22} +  
  17 N_{23} - N_{24} + 2 N_{25} + 
  N_{26} + 3 N_{27}, \cr
   & \cr
  0=&  -5 N_4 + 3  N_5 -2  N_6 + N_7 -2  N_8 +  N_9 -
    N_{10} -2  N_{12} + N_{13}- 
    N_{14} + 3 N_{16} - 2 N_{17} +\cr
  &  N_{18} -  N_{19}+
   3 N_{20} -2  N_{21} +  N_{22} -  N_{23} +3  N_{24} - 2 N_{25} + 
  N_{26} - N_{27}, \cr
   & \cr
  0= & - N_1 -  N_2 - 2 N_3 - N_5 -  N_6 -2  N_7 +  N_8  +
     N_{10} +  N_{11} -  N_{16} - 
     N_{17} - 2 N_{18} -3  N_{19}- \cr
   & N_{20}  -  N_{22} - N_{23}  +  N_{25} + 
  N_{26} + 2 N_{27}, \cr
   &\cr
  0= &-N_0  -  N_2  - N_3 -  N_4  - N_6 -  N_7  - N_8 +  N_9 +
      N_{11}   -  N_{17}  - N_{18} -2  N_{19}+
   N_{20}  -\cr
   & N_{21} - N_{23} + N_{24}  + 
  N_{26} +  N_{27}. \cr }}
The trivial solution with just the D-branes on top of the orientifold
plane corresponds to $N_{15}=4$ with all other CP-factors vanishing. 

\vfill\eject 
\vskip 0.8cm
\vbox{ \centerline{\vbox{ \hbox{\vbox{\offinterlineskip
\def\tablespace{height2pt&\omit&&\omit&&\omit&&
 \omit&\cr}
\def\tablerule{\tablespace\noalign{\hrule}\tablespace}

\hrule\halign{&\vrule#&\strut\hskip0.2cm\hfill #\hfill\hskip0.2cm\cr
& $(L_1,L_2,L_3,L_4,L_5)$ &&  degeneracy  &&  CP-factor && moduli $n_{(S,A)}$ &\cr
\tablerule
& $(0,0,0,0,0)$ && $1$ && $N_0$ && $0$ &\cr
\tablerule
& $(0,2,0,0,0)$ && $3$ && $N_1$ && $2_{(2,0)}$ &\cr
\tablerule
& $(0,2,2,0,0)$ && $3$ && $N_2$ && $5_{(4,1)}$ &\cr
\tablerule
& $(0,2,2,2,0)$ && $1$ && $N_3$ && $12_{(9,3)}$ &\cr
\tablerule
& $(0, 0, 0, 0, 2)$ && $1$ && $N_4$ && $0$ &\cr
\tablerule
& $(0,2,0,0,2)$ && $3$ && $N_5$ && $4_{(4,0)}$ &\cr
\tablerule
& $(0,2,2,0,2)$ && $3$ && $N_6$ && $11_{(8,3)}$ &\cr
\tablerule
& $(0,2,2,2,2)$ && $1$ && $N_7$ && $24_{(15,9)}$ &\cr
\tablerule
& $(0, 0, 0, 0, 4)$ && $1$ && $N_8$ && $3_{(3,0)}$ &\cr
\tablerule
& $(0, 2, 0, 0, 4)$ && $3$ && $N_9$ && $8_{(7,1)}$ &\cr
\tablerule
& $(0,2,2,0,4)$ && $3$ && $N_{10}$ && $18_{(13,5)}$ &\cr
\tablerule
& $(0,2,2,2,4)$ && $1$ && $N_{11}$ && $37_{(24,13)}$ &\cr
\tablerule
& $(0, 0, 0, 0, 6)$ && $1$ && $N_{12}$ && $7_{(4,3)}$ &\cr
\tablerule
& $(0, 2, 0, 0, 6)$ && $3$ && $N_{13}$ && $15_{(9,6)}$ &\cr
\tablerule
& $(0, 2, 2, 0, 6)$ && $3$ && $N_{14}$ && $31_{(17,14)}$ &\cr
\tablerule
& $(0, 2, 2, 2, 6)$ && $1$ && $N_{15}$ && $63_{(32,31)}$ &\cr
\tablerule
& $(0, 0, 0, 0, 8)$ && $1$ && $N_{16}$ && $4_{(4,0)}$ &\cr
\tablerule
& $(0, 2, 0, 0, 8)$ && $3$ && $N_{17}$ && $11_{(8,3)}$ &\cr
\tablerule
& $(0, 2, 2, 0, 8)$ && $3$ && $N_{18}$ && $24_{(15,9)}$ &\cr
\tablerule
& $(0, 2, 2, 2, 8)$ && $1$ && $N_{19}$ && $50_{(28,22)}$ &\cr
\tablerule
& $(0, 0, 0, 0, 10)$ && $1$ && $N_{20}$ && $3_{(3,0)}$ &\cr
\tablerule
& $(0, 2, 0, 0, 10)$ && $3$ && $N_{21}$ && $8_{(7,1)}$ &\cr
\tablerule
& $(0, 2, 2, 0, 10)$ && $3$ && $N_{22}$ && $18_{(13,5)}$ &\cr
\tablerule
& $(0, 2, 2, 2, 10)$ && $1$ && $N_{23}$ && $37_{(24,13)}$ &\cr
\tablerule
& $(0, 0, 0, 0, 12)$ && $1$ && $N_{24}$ && $0$ &\cr
\tablerule
& $(0, 2, 0, 0, 12)$ && $3$ && $N_{25}$ && $2_{(2,0)}$ &\cr
\tablerule
& $(0, 2, 2, 0, 12)$ && $3$ && $N_{26}$ && $5_{(4,1)}$ &\cr
\tablerule
& $(0, 2, 2, 2, 12)$ && $1$ && $N_{27}$ && $12_{(9,3)}$ &\cr
}\hrule}}}} 
\centerline{ \hbox{{\bf
Table 6:}{\it ~~ boundary states for the (1,3,3,3,13) Gepner model}}} } 
\vskip 0.5cm
\noindent

\vfill\eject
\listrefs

\bye
\end